\documentclass[journal]{IEEEtran}
\usepackage{cite,graphicx,amsmath,psfrag,bm}
\usepackage[usenames,dvipsnames]{xcolor}
\usepackage{mathabx}
\usepackage{subfig}
\usepackage{cancel}
\usepackage{epstopdf}
\usepackage{pstool}
\usepackage{color}
\usepackage{soul}

\newcommand{\ve}[1]{\boldsymbol{#1}}
\newcommand{\te}[1]{\overline{\overline{#1}}}

\newcounter{tempEquationCounter}
\newcounter{thisEquationNumber}
\newenvironment{floatEq}
{\setcounter{thisEquationNumber}{\value{equation}}\addtocounter{equation}{1}
\begin{figure*}[!t]
\normalsize\setcounter{tempEquationCounter}{\value{equation}}
\setcounter{equation}{\value{thisEquationNumber}}
}
{\setcounter{equation}{\value{tempEquationCounter}}
\hrulefill\vspace*{4pt}
\end{figure*}
}

\begin{document}

\title{General Metasurface Synthesis\\ Based on Susceptibility Tensors}

\author{Karim~Achouri,
        Mohamed~A.~Salem,
        and~Christophe~Caloz,~\IEEEmembership{Fellow,~IEEE}

\thanks{K. Achouri, M. A. Salem and C. Caloz are with the Department
of Electrical Engineering, $\acute{\mathrm{E}}$cole polytechnique de Montr$\acute{\mathrm{e}}$al, Montr$\acute{\mathrm{e}}$al,
QC, H3T 1J4 Canada (e-mail: karim.achouri@polymtl.ca).}
\thanks{Manuscript received MONTH XX, 2014; revised MONTH XX, 2014.}}

\markboth{IEEE Transactions on Antennas and Propagation,~Vol.~X, No.~Y, Month~Z}%
{Shell \MakeLowercase{\textit{et al.}}: Metasurface Synthesis Using Reduced Susceptibility Tensors}

\maketitle

\begin{abstract}
A general method, based on susceptibility tensors, is proposed for the synthesis of metasurfaces transforming arbitrary incident waves into arbitrary reflected and transmitted waves. The proposed method exhibits two advantages: 1)~it is inherently vectorial, and therefore better suited for full vectorial (beyond paraxial) electromagnetic problems, 2)~it provides closed-form solutions, and is therefore extremely fast. Incidentally, the method reveals that a metasurface is fundamentally capable to transform up to four independent wave triplets (incident, reflected and refracted waves). In addition, the paper provides the closed-form expressions relating the synthesized susceptibilities and the scattering parameters simulated within periodic boundary conditions, which allows one to design the scattering particles realizing the desired susceptibilities. The versatility of the method is illustrated by examples of metasurfaces achieving the following transformations: generalized refraction, reciprocal and non-reciprocal polarization rotation, Bessel vortex beam generation, and orbital angular momentum multiplexing.
\end{abstract}


\IEEEpeerreviewmaketitle

\section{Introduction}

Metasurfaces~\cite{kuester2003av,Holloway2009,holloway2012overview} are dimensional reductions of volume metamaterials~\cite{capolino2009theory,engheta2006metamaterials,caloz2005electromagnetic} and functional extensions of frequency selective surfaces~\cite{MunkFSS}. They are composed of two-dimensional arrays of sub-wavelength scattering particles engineered in such a manner that they transform incident waves into desired reflected and transmitted waves. Compared to volume metamaterials, metasurfaces offer the advantage of being lighter, easier to fabricate and less lossy due to their reduced dimensionality, while compared to frequency selective surfaces, they provide greater flexibility and functionalities.

A myriad of metasurfaces have been reported in the literature. For instance, one may mention metasurfaces providing tunable reflection and transmission coefficients~\cite{holloway2005reflection}, plane wave refraction~\cite{Pfeiffer2013a}, single-layer perfect absorption~\cite{Ra2013}, polarization twisting~\cite{shi2014dual}, and vortex wave generation~\cite{capasso1}, and many more metasurface structures and applications are expected to emerge in coming years.

To date, only a small number of metasurface synthesis techniques have been reported~\cite{PhysRevApplied.2.044011,6905746,6477089,Salem2013c}. The first two, \cite{PhysRevApplied.2.044011} and \cite{6905746}, are based on impedance tensors corresponding to the relations~\eqref{eq:InvProb}. They are similar to the case presented in Sec.~\ref{sec:single_transf}, which represents only a particular case of the general method proposed in the paper. The third technique, \cite{6477089}, relates the waves reflected and transmitted from the metasurface to the polarizabilities of a single scattering element in the case of a normally incident plane wave. In contrast, this paper deals with wave of arbitrary incident angle and arbitrary type. Finally, the last paper, \cite{Salem2013c}, proposes a technique called the momentum transformation method, which is a spectral ($\ve{k}$) method, that is particularly suitable for paraxial wave problems. It can also handle full vectorial problems but this involves extra complexity compared to the scalar case.

We propose here an alternative synthesis method, which is inherently vectorial and which may therefore be easier to implement in full vectorial electromagnetic situations. Moreover, this method leads to \emph{closed-form} solutions, and is hence extremely fast. It describes the metasurface in terms of surface susceptibility tensors in the space domain, where the susceptibility tensors are related through Generalized Sheet Transition Conditions (GSTCs)~\cite{kuester2003av} to the incident, reflected and transmitted fields around the structure.

The paper is organized as follows. Section~\ref{sec:metasurf_synth_prob} defines the synthesis problem to be solved. Section~\ref{sec:MS_boundary_cond} explains why conventional textbook boundary conditions are not adequate to handle metasurface problems and establishes, with the help of the Appendix, the metasurface GSTCs for bi-anisotropic metasurfaces. Section~\ref{sec:synthesis} is the core of paper. It presents the proposed synthesis method, points out the fundamental possibility for a metasurface to simultaneously handle several independent waves and derives closed-form expressions for its susceptibilities for the case of one and two wave transformations. The versatility of the method is illustrated in Sec.~\ref{sec:examples} by various examples, where the metasurface is synthesized so as to provide generalized refraction, reciprocal and non-reciprocal polarization rotation, Bessel vortex beam generation, and orbital angular momentum multiplexing. Finally, conclusions are provided in Sec.~\ref{sec:conc}.

\section{Metasurface Synthesis Problem}
\label{sec:metasurf_synth_prob}

A metasurface is an electromagnetic two-dimensional structure with sub-wavelength thickness ($\delta\ll\lambda$). The metasurface may be finite, with dimensions $L_x\times{L_y}$, or infinite. It is typically composed of a non-uniform arrangement of planar scattering particles (full or slotted patches, straight or curved strips, various types of crosses, etc.) that transforms incident waves into specified reflected and transmitted waves.

\begin{figure}[h!]
\centering
\includegraphics[width=0.75\linewidth]{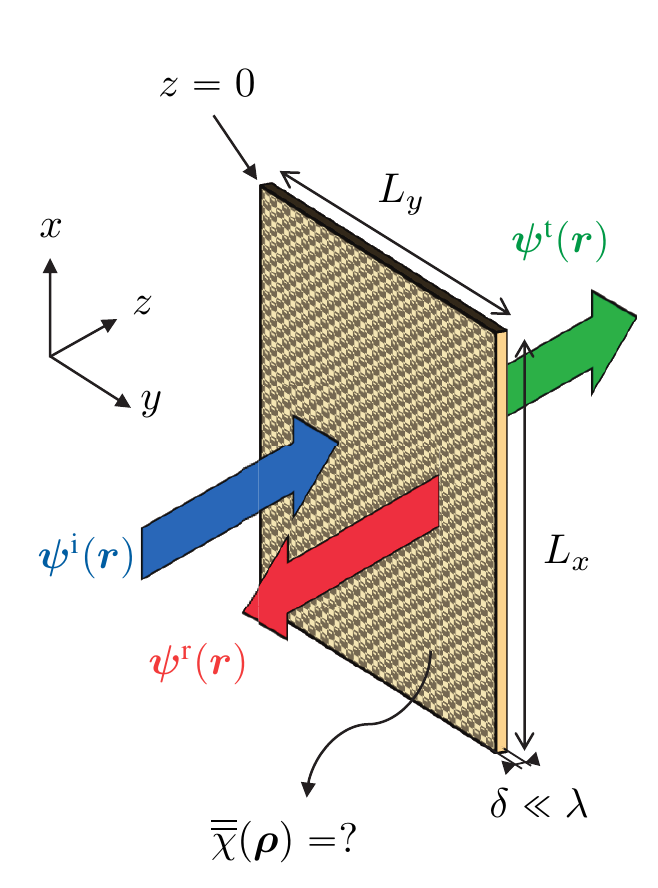}
\caption{Metasurface synthesis (inverse) problem to solve. A metasurface, generally defined as an electromagnetic two-dimensional non-uniform structure of extent $L_x\times{L_y}$ with sub-wavelength thickness ($\delta\ll\lambda$), is placed at $z=0$. Determine the surface susceptibility tensors $\ve{\chi}(\ve{\rho})$ of the metasurface transforming an arbitrary specified incident wave $\ve{\psi}^\text{i}(\ve{r})$ into an arbitrary specified reflected wave $\ve{\psi}^\text{r}(\ve{r})$ and an arbitrary specified transmitted wave $\ve{\psi}^\text{t}(\ve{r})$.}
\label{fig:MS}
\end{figure}

Figure~\ref{fig:MS} shows the synthesis problem to solve. How can one synthesize a metasurface that transforms an arbitrary specified incident wave, $\ve{\psi}^\text{i}(\ve{r})$, into an arbitrary specified reflected wave, $\ve{\psi}^\text{r}(\ve{r})$, and an arbitrary specified transmitted wave, $\ve{\psi}^\text{t}(\ve{r})$, assuming monochromatic waves? Here the solution will be expressed in terms of the transverse susceptibility tensor functions of $\ve{\rho}=x\ve{\hat{x}}+y\ve{\hat{y}}$, $\te{\chi}_\text{ee}(\ve{\rho})$, $\te{\chi}_\text{mm}(\ve{\rho})$, $\te{\chi}_\text{em}(\ve{\rho})$ and $\te{\chi}_\text{me}(\ve{\rho})$, which represent the electric/magnetic (e/m) transverse polarization responses (first subscript) to transverse electric/magnetic (e/m) field excitations (second subscript).

The synthesis procedure will always yield $\te{\chi}_\text{ee}(\ve{\rho})$, $\te{\chi}_\text{mm}(\ve{\rho})$, $\te{\chi}_\text{em}(\ve{\rho})$ and $\te{\chi}_\text{me}(\ve{\rho})$ results, but will not guarantee that these results can be practically implemented using planar scattering particles. For instance, if the susceptibilites exhibit multiple spatial variations per wavelength, it may be difficult or impossible to realize. In such cases, one has to determine whether some features may be neglected or one may have to relax the design constraints (e.g. allow higher reflection or increase the metasurface dimensions).

The complete synthesis of a metasurface typically consists in two steps: 1)~determination of the mathematical transfer function of the metasurface producing the specified fields, which is generally a continuous function of the transverse dimensions of the metasurface; 2)~discretization of the transfer function obtained in 1) according to a two-dimensional lattice and determination of the scattering particles realizing the corresponding transfer function at each lattice site. Step~2) involves a full-wave parametric analysis of judiciously selected scattering particles, from which magnitude and phase maps are established to find the appropriate particle geometries for building the metasurface using the periodic boundary condition approximation~\cite{capasso1}. Since this second step involves scattering parameters, the paper also provides transformation formulas between susceptibilities and scattering parameters to enable the complete synthesis of the metasurface.

\section{Metasurface Boundary Conditions in Terms of Surface Susceptibility Tensors}
\label{sec:MS_boundary_cond}
A metasurface may be considered as an electromagnetic \emph{discontinuity} in space. Conventional textbook electromagnetic boundary conditions do not apply to such a discontinuity. As was  pointed out by Schelkunoff~\cite{Schelkunoff1}, the mathematical formulation of the conventional boundary conditions is not rigorous in the case of field discontinuities caused by sources, such as surface charges and currents, although it yields satisfactory results away from the discontinuities. Assuming an interface at $z=0$, the conventional boundary conditions relate the fields at $z=0^\pm$, but fail to describe the field behavior at the discontinuity itself ($z = 0$). This discrepancy is due to the fact that Stokes and Gauss theorems used to derive them assume field continuity in all the regions they apply to, including the interface, whereas the fields may be discontinuous due to the presence of sources. For instance, consider the conventional boundary condition for the normal component of the displacement vector $\ve{D}$ in the presence of surface charges $\rho_s$,
\begin{equation}
\label{eq:BC_D_rho}
\ve{\hat{z}}\cdot\left.\ve{D}\right|_{z=0^-}^{0^+}
=\rho_s.
\end{equation}
This relation is derived by applying Gauss theorem, \mbox{$\iiint_V \nabla \cdot \mathbf{D} dV = \oiint_S \mathbf{D} \cdot \hat{\mathbf{n}} dS$}, to a volume $V$ enclosed by the surface $S$ including the interface discontinuity with $\hat{\mathbf{n}}$ the normal unit vector to $S$. This theorem rigorously applies only if $\mathbf{D}$ is continuous inside the entire volume $V$, whereas in the case of a discontinuous $\mathbf{D}$, its projection onto $S$ is not defined at the interface and application of this theorem is not rigorously correct. Thus, since a metasurface may be modeled by Huygens sources~\cite{GrbicLightBending}, the correct field behavior on the metasurface cannot be determined using the conventional boundary conditions and rigorous boundary conditions, namely GSTCs, must be applied, as will be done next. It should be noted that, from a  physical perspective, a metasurface structure is not a single interface but rather a thin inhomogeneous slab, and may be naturally treated as such. However, it is much simpler to treat the metasurface as a single interface using rigorous GSTCs, which is allowed by the fact that it is electromagnetically thin.

Rigorous GSTCs, treating discontinuities in the sense of distributions, were derived by Idemen~\cite{Idemen1973}. The corresponding relations pertaining to this work, first applied by Kuester {\it et al.} to metasurfaces~\cite{kuester2003av}, are derived in Appendix~\ref{Appendix} for the sake of clarity and completeness. They may be written as\footnote{Throughout the paper, the medium surrounding the metasurface is assumed to be vacuum, with permittivity and permeability $\epsilon_0$ and $\mu_0$, respectively. Therefore, for notational compactness, and also to avoid confusion between the subscript `0' meaning `vacuum' or meaning `first order discontinuity' (Appendix~\ref{Appendix}), the subscript `0' for `vacuum' is suppressed everywhere.}
\begin{subequations}
\label{eq:BC}
\begin{align}
\hat{z}\times\Delta\ve{H}
&=j\omega\ve{P}_{\parallel}-\hat{z}\times\nabla_{\parallel}M_{z},\label{eq:CurlH}\\
\Delta\ve{E}\times\hat{z}
&=j\omega\mu \ve{M}_{\parallel}-\nabla_{\parallel}\bigg(\frac{P_{z}}{\epsilon }\bigg)\times\hat{z},\label{eq:CurlE}\\
\hat{z}\cdot\Delta\ve{D}
&=-\nabla\cdot\ve{P}_{\parallel},\label{eq:divD}\\
\hat{z}\cdot\Delta\ve{B}
&=-\mu \nabla\cdot\ve{M}_{\parallel}.
\end{align}
\end{subequations}
\noindent In these relations, the terms in the left-hand sides represent the differences between the fields on the two sides of the metasurface, whose cartesian components are defined as
\begin{equation}
\label{eq:field_diff}
\Delta \Psi_{u}
=\ve{\hat{u}}\cdot\Delta\ve{\Psi}(\ve{\rho})\Bigr|_{z=0^{-}}^{0^{+}}
=\Psi_{u}^\text{t}-(\Psi_{u}^\text{i}+\Psi_{u}^\text{r}),\; u=x,y,z,
\end{equation}
where $\ve{\Psi}(\ve{\rho})$ represents any of the fields $\ve{H}$, $\ve{E}$, $\ve{D}$ or $\ve{B}$, and where the superscripts i, r, and t denote incident, reflected and transmitted fields, and $\ve{P}$ and $\ve{M}$ are the electric and magnetic surface polarization densities, respectively. In the most general case of a bi-anisotropic medium, these densities are related to the acting (or local) fields, $\ve{E}_\text{act}$ and $\ve{H}_\text{act}$, by~\cite{kong1986electromagnetic,lindell1994electromagnetic}
\begin{subequations}
\label{eq:pola_dens}
\begin{align}
\ve{P}&=\epsilon N\langle\te{\alpha}_{\text{ee}}\rangle\ve{E}_\text{act}+\sqrt{\mu \epsilon } N\langle\te{\alpha}_{\text{em}}\rangle\ve{H}_\text{act},\\
\ve{M}&=N\langle\te{\alpha}_{\text{mm}}\rangle\ve{H}_\text{act}+\sqrt{\frac{\epsilon }{\mu }}N\langle\te{\alpha}_{\text{mm}}\rangle\ve{E}_\text{act},
\end{align}
\end{subequations}
\noindent where the $\langle\te{\alpha}_{\text{ab}}\rangle$ terms represent the averaged polarizabilities of a given scatterer, and $N$ is the number of scatterers per unit area. The acting fields are, by definition, the average fields on both sides of the surface taking into account the contributions of all the scattering particles (coupling effects) except that of the particle being considered. The contribution of this particle may be modeled by replacing it with a disk of radius $R$ encompassing its electric and magnetic current dipoles. Kuester {\it et al.} express the fields of this disk as functions of the polarization densities $\ve{P}$ and $\ve{M}$~\cite{kuester2003av}, from which relations \eqref{eq:pola_dens} can be rewritten as functions of the average fields. Their relations, with averaged polarizabilities replaced by surface susceptibilities for macroscopic description, read
\begin{subequations}
\label{eq:pola_dens_2}
\begin{align}
 \ve{P}&=\epsilon \te{\chi}_\text{ee}\ve{E}_\text{av}+\te{\chi}_\text{em}\sqrt{\mu \epsilon }\ve{H}_\text{av},\\
 \ve{M}&=\te{\chi}_\text{mm}\ve{H}_\text{av}+\te{\chi}_\text{me}\sqrt{\frac{\epsilon }{\mu }}\ve{E}_\text{av},
\end{align}
\end{subequations}
where the average fields are defined as
\begin{equation}
\label{eq:field_av}
\Psi_{u,\text{av}}
=\ve{\hat{u}}\cdot\ve{\Psi}_\text{av}(\ve{\rho})
=\frac{\Psi_u^\text{t}+(\Psi_u^\text{i}+\Psi_u^\text{r})}{2},
\; u=x,y,z,
\end{equation}
\noindent where $\ve{\Psi}(\ve{\rho})$ represents either $\ve{H}$ or $\ve{E}$. The utilization of surface susceptibilities, which represent the actual macroscopic quantities of interest, allows for an easier description of the metasurface than particle averaged polarizabilities and densities in~\eqref{eq:pola_dens}.

The surface may be infinite or finite with dimensions $L_x\times{L_y}$. The two problems are automatically solved by specifying the fields $\Psi_u^\text{i}$, $\Psi_u^\text{r}$ and $\Psi_u^\text{t}$ in~\eqref{eq:field_diff} and~\eqref{eq:field_av} to be of infinite or finite $L_x\times{L_y}$ extent in the former and latter cases, respectively. In the finite case, truncation practically corresponds to placing a sheet of absorbing material around the metasurface. This operation neglects diffraction at the edges of the metasurface, as is safely allowed by the fact that a metasurface is generally electrically very large, but properly accounts for the finiteness of the aperture via the GSTCs~\eqref{eq:CurlH} and \eqref{eq:CurlE}.

\section{Synthesis Method}
\label{sec:synthesis}

\subsection{Assumptions}\label{sec:assumptions}
The proposed synthesis method solves the inverse problem depicted in Fig.~\ref{fig:MS}, where the electromagnetic fields are specified everywhere (for all $\ve{\rho}$) in the $z=0$ plane on both sides of the metasurface and the properties of the metasurface are the unknowns to be determined. We specifically aim at finding the susceptibilities that transform specified incident waves into specified transmitted and reflected waves. The method essentially consists in solving Eqs.~\eqref{eq:BC} for the components of the susceptibility tensors in~\eqref{eq:pola_dens_2}.

The last terms in~\eqref{eq:CurlH} and \eqref{eq:CurlE} involve the transverse derivatives of the normal components of the polarization densities, namely $\nabla_\parallel M_z$ and $\nabla_\parallel P_z$. Solving the inverse problem for non-zero $M_{z}$ and/or $P_{z}$ would be quite involved since this would require solving the set of coupled non-homogenous partial differential equations formed by \eqref{eq:CurlH} and \eqref{eq:CurlE} with nonzero $\nabla_\parallel M_z$ and $\nabla_\parallel P_z$. Although such a problem could be generally addressed by means of numerical analysis, we enforce here $P_{z}=M_{z}=0$, which will lead to convenient closed-form solutions for the susceptibilities\footnote{This restriction may limit the physical realizability of the metasurface in some cases, in the sense that the corresponding synthesized susceptibilities might be excessively difficult to realize with practical scattering particles. In such cases, the restriction might be removed without changing the main spirit of the method but at the risk of losing the closed-form nature of the solution.}. As shall be seen next, this restriction still allows the metasurface to realize a large number of operations, given the large number of degrees of freedom provided by combinations of its bi-anisotropic susceptibility tensor components.

The method needs considering only \eqref{eq:CurlH} and \eqref{eq:CurlE} as these two equations involve all the transverse field components, which is sufficient to completely describe the fields at each side of the metasurface according to the uniqueness theorem. These two equations, with $P_{z}=M_{z}=0$, represent a set a four linear equations relating the transverse electric and magnetic fields to the effective surface susceptibilities. Thus, the solution of the inverse problem will consist in determining four transverse effective susceptibility tensors in~\eqref{eq:pola_dens_2}.

However, the proposed method can also handle the most general case including normal components of the polarization densities, which may be of practical interest (e.g. conducting rings in the metasurface plane, producing $M_z$ contributions). To show this, Sec.~\ref{sec:normal_synthesis} presents the developments required to synthesize metasurfaces involving nonzero $P_{z}$ and $M_{z}$.

\subsection{General Solution for Surface Susceptibilities}
\label{sec:gen_sol}

As announced in Sec.~\ref{sec:assumptions}, the four susceptibility tensors in~\eqref{eq:pola_dens_2} are restricted to their four transverse components, and these components will be determined for the specified fields using~\eqref{eq:CurlH} and \eqref{eq:CurlE} with $P_{z}=M_{z}=0$, i.e., using the notation in~\eqref{eq:field_diff} and~\eqref{eq:field_av},
\begin{subequations}
\label{eq:InvProb}
\begin{multline}
\label{eq:diffH}
\binom{-\Delta H_y}{\Delta H_x}=j\omega\epsilon  \begin{pmatrix} \chi_{\text{ee}}^{xx} & \chi_{\text{ee}}^{xy} \\ \chi_{\text{ee}}^{yx} & \chi_{\text{ee}}^{yy} \end{pmatrix}\binom{E_{x,\text{av}}}{E_{y,\text{av}}}\\+j\omega\sqrt{\epsilon \mu }\begin{pmatrix} \chi_{\text{em}}^{xx} & \chi_{\text{em}}^{xy} \\ \chi_{\text{em}}^{yx} & \chi_{\text{em}}^{yy} \end{pmatrix}\binom{H_{x,\text{av}}}{H_{y,\text{av}}},
\end{multline}

\begin{multline}
\label{eq:diffE}
\binom{\Delta E_y}{-\Delta E_x}=j\omega\mu  \begin{pmatrix} \chi_{\text{mm}}^{xx} & \chi_{\text{mm}}^{xy} \\ \chi_{\text{mm}}^{yx} & \chi_{\text{mm}}^{yy} \end{pmatrix}\binom{H_{x,\text{av}}}{H_{y,\text{av}}}\\+j\omega\sqrt{\epsilon \mu }\begin{pmatrix} \chi_{\text{me}}^{xx} & \chi_{\text{me}}^{xy} \\ \chi_{\text{me}}^{yx} & \chi_{\text{me}}^{yy} \end{pmatrix}\binom{E_{x,\text{av}}}{E_{y,\text{av}}}.
\end{multline}
\end{subequations}

Assuming single incident, reflected and transmitted waves (only one wave of each of the three types), the system~\eqref{eq:InvProb} contains 4 equations for a total number of 16 unknown susceptibility components. It is thus \emph{underdetermined} as such, and it can be solved only by restricting the number of \emph{independent} susceptibilities to 4. This single-transformation underdetermination of~\eqref{eq:InvProb} reveals two important facts: i)~\emph{Many different combinations of susceptibilities produce the same fields}; ii)~A metasurface has the \emph{fundamental capability to simultaneously manipulate several linearily independent incident, reflected and transmitted waves}. Specifically, a metasurface, as defined by~\eqref{eq:diffE}, can in principle manipulate \emph{up to 4 sets of incident, reflected and transmitted waves}. If $T$ ($T=1,2,3,4$) waves are to be manipulated, corresponding to $4T$ independent equations obtained by writing the 4 equations in~\eqref{eq:diffE} for each of the field sets $\ve{\Psi}_n(\ve{\rho})$ ($n=1,\ldots,T$, $\ve{\Psi}$ representing either $\ve{E}$ or $\ve{H}$), $4T$ (4,8,12,16) susceptibilities have to be specified.

Two approaches may be considered to reduce the number of independent unknown susceptibilities when $T<4$. A first approach could consist in using more than $4T$ (4,8,12) susceptibilities but enforcing relationships between some of them to ensure a maximum of $4T$ independent unknowns. For example, the conditions of reciprocity and losslessness would be a possible way to link some susceptibilities together, if this is compatible with design specifications. According to Kong \cite{kong1986electromagnetic} and Lindell~\cite{lindell1994electromagnetic}, the conditions for reciprocity and losslessness are
\begin{subequations}
\label{eq:reciprocityLoss}
\begin{equation}
\label{eq:reciprocity}
\te{\chi}_{\text{ee}}^{\text{T}} = \te{\chi}_{\text{ee}}, \quad \te{\chi}_{\text{mm}}^{\text{T}}=\te{\chi}_{\text{mm}}, \quad \te{\chi}_{\text{me}}^{\text{T}}=-\te{\chi}_{\text{em}},
\end{equation}
\begin{equation}
\label{eq:lossless}
\te{\chi}_{\text{ee}}^{\text{T}} = \te{\chi}_{\text{ee}}^*, \quad \te{\chi}_{\text{mm}}^{\text{T}}=\te{\chi}_{\text{mm}}^*, \quad \te{\chi}_{\text{me}}^{\text{T}}=\te{\chi}_{\text{em}}^*,
\end{equation}
\end{subequations}
\noindent respectively, where the superscripts T and $*$ denote the matrix transpose and complex conjugate operations, respectively. Enforcing conditions between susceptibilities also enforces conditions on the fields on both sides of the metasurface. Therefore, this approach restricts the diversity of electromagnetic transformations achievable with the metasurface.

A second approach, providing a more general synthesis method for quasi-arbitrary electromagnetic transformations, is then preferred. This approach consists in selecting only $T$ susceptibility tensor components in each of the $4T$ equations included in~\eqref{eq:InvProb}. The number of possible susceptibilities in each equation is given by $C_k^n=\frac{n!}{k!(n-k)!}$, where $n=4$ and $k=T$. Therefore, for $T=1$, $C_k^n=C_1^4=4$ and the total number of different combinations for the $4$ equations is $(C_1^4)^4=4^4=256$. By the same token, we respectively have for $T=2,3$ and $4$ the following total number of different combinations $(C_2^4)^4=6^4=1296, (C_3^4)^4=4^4=256$ and $(C_4^4)^4=1$. Among these combinations, those that do not satisfy~\eqref{eq:reciprocityLoss} naturally correspond to non-reciprocal and/or lossy designs. It is obviously impossible to cover these huge numbers of synthesis combination sets for $T<4$. We will therefore next, without loss of generality, restrict our attention to the cases of single (Sec.~\ref{sec:single_transf}) and double (Sec.~\ref{sec:double_transf}) transformations with mono-anisotropic metasurfaces. The solutions for bi-anisotropic, triple-wave and quadruple-wave metasurfaces can be obtained by following exactly the same procedure.

Note that these considerations hold in the most general case of transformations where the amplitude, phase and polarization of the incident field are all modified by the metasurface. Under such conditions, the system in~\eqref{eq:InvProb} can indeed handle up to $T=4$ independent wave triplets. In the particular case of single-triplet transformation, only 4~susceptibilites (2 electric and 2 magnetic) are generally required, as will be shown shortly. However, as will be seen in the example of generalized refraction in Sec.~\ref{sec:Refrac}, depending on the transformation and choice of susceptibilities, only 2~susceptibilities (1 electric and 1 magnetic) may be sufficient. In general, this factor 2 reduction in required susceptibilities indicates that the number of possible transformations doubles, i.e. $T'=2T$.

What has been described so far in this section represents the first step of the synthesis procedure. As mentioned in Sec.~\ref{sec:metasurf_synth_prob}, the second step consists in determining the scattering particles realizing the transfer function corresponding to the synthesized susceptibilities. In this second step, one computes the full-wave scattering parameters for an isolated unit cell within 2D periodic boundary conditions, where periodicity is an approximation of typically slowly varying scattering elements in the plane of the metasurface~\cite{GrbicLightBending,asadchy2011simulation,asadchy2014determining,Pfeiffer2013a}. The periodic boundary conditions in full-wave analysis are generally restricted to rectilinearly propagating waves. Now, the prescribed waves may change directions at the metasurface (e.g. case of generalized refraction, Sec.~\ref{sec:Refrac}). In such cases, ``rectilinear'' periodic boundary conditions cannot directly describe the physics of the problem. However, the results they provide correspond to a rigorous mapping with the physical problem, and they may thus be rigorously used in the synthesis, as will be illustrated in Sec.~\ref{sec:Refrac}.
\subsection{Single Transformation}
\label{sec:single_transf}
We consider here the problem of single ($T=1$) transformation [only one specified wave triplet: ($\ve{\Psi}^\text{i}$, $\ve{\Psi}^\text{r}$, $\ve{\Psi}^\text{t}$)] for a mono-anisotropic ($\te{\chi}_\text{em}\equiv\te{\chi}_\text{me}=0$) and uniaxial ($\chi_\text{ee}^{xy}\equiv\chi_\text{ee}^{yx}\equiv\chi_\text{mm}^{xy}\equiv\chi_\text{mm}^{yx}=0$), and hence non-gyrotropic and reciprocal, metasurface. Solving~\eqref{eq:InvProb} under these conditions yields the following simple relations for the remaining 4 susceptibilities
\begin{subequations}
\label{eq:chi_diag}
\begin{align}
\chi_{\text{ee}}^{xx}&=\frac{-\Delta H_{y}}{j\omega\epsilon  E_{x,\text{av}}},\label{eq:chi_diag_Exx}\\
\chi_{\text{ee}}^{yy}&=\frac{\Delta H_{x}}{j\omega\epsilon  E_{y,\text{av}}},\label{eq:chi_diag_Eyy}\\
\chi_{\text{mm}}^{xx}&=\frac{\Delta E_{y}}{j\omega\mu  H_{x,\text{av}}},\label{eq:chi_diag_Mxx}\\
\chi_{\text{mm}}^{yy}&=\frac{-\Delta E_{x}}{j\omega\mu  H_{y,\text{av}}},\label{eq:chi_diag_Myy}
\end{align}
\end{subequations}
where, according to~\eqref{eq:field_diff} and~\eqref{eq:field_av}, $\Delta H_y=H_y^\text{t}-(H_y^\text{i}+H_y^\text{r})$, $E_{x,\text{av}}=\frac{E_x^\text{t}+(E_x^\text{i}+E_x^\text{r})}{2}$, and so on.

By synthesis, a metasurface with the susceptibilities given by~\eqref{eq:chi_diag} will produce exactly the specified reflected and transmitted transverse components of the fields when the metasurface is illuminated by the specified incident field. Since the longitudinal fields are completely determined from the transverse components, according to the uniqueness theorem, the complete specified electromagnetic fields are exactly generated by the metasurface.

Consistency with Maxwell equations can be easily verified. Consider for instance \eqref{eq:divD} along with the relation~\mbox{$\ve{D}=\epsilon\ve{E}+\ve{P}$},
\begin{equation}
\begin{split}
D_{z}\Bigr|_{z=0^{-}}^{0^{+}}&=\epsilon E_{z}\Bigr|_{z=0^{-}}^{0^{+}}+ P_{z}
=\epsilon \Delta E_{z}+ P_{z}\\
&=-\nabla\cdot\ve{P}_{\perp}.
\end{split}
\end{equation}

Substituting in this relation the relations \eqref{eq:pola_dens_2} for $\ve{P}_{\perp}$ and remembering our assumption that $P_z=0$ (Sec.~\ref{sec:assumptions}), we find
\begin{equation}
\label{eq:DeltaEz1}
\Delta E_{z}=-\frac{\partial}{\partial x}\left(\chi_{\text{ee}}^{xx}E_{x,\text{av}}\right)-\frac{\partial}{\partial y}\left(\chi_{\text{ee}}^{yy}E_{y,\text{av}}\right),
\end{equation}
\noindent which upon substitution of~\eqref{eq:chi_diag} becomes
\begin{equation}
\begin{split}
\label{eq:DeltaEz2}
\Delta E_{z}&=E_z^\text{t}-E_z^\text{i}-E_z^\text{r}\\
&=\frac{j}{\omega\epsilon}\left[\frac{\partial}{\partial y}(H_x^\text{t} -H_x^\text{i}-H_x^\text{r})-\frac{\partial}{\partial x}(H_y^\text{t} -H_y^\text{i}-H_y^\text{r})\right].
\end{split}
\end{equation}

\noindent This equation represents a relation between linear combinations of the longitudinal electric fields and derivatives of the transverse magnetic fields of the incident (k$=$i), reflected (k$=$r) and transmitted (k$=$t) waves. From linearity, and subsequent superposition, these equations may be decomposed as
\begin{equation}
\label{eq:Ez_max}
E_z^\text{k}=\frac{j}{\omega\epsilon}\left(\frac{\partial{H_x^\text{k}}}{\partial{y}}-\frac{\partial{H_y^\text{k}}}{\partial{x}}\right)
=\frac{j}{\omega\epsilon}\left(\nabla_t\times\ve{H}^\text{k}\right)_z,
\end{equation}
which is nothing but the projection of Maxwell-Amp\`ere equation upon the $z$ direction. This shows that the longitudinal fields are well defined with the relations \eqref{eq:chi_diag} in accordance with the uniqueness theorem.

We are now interested in finding the relationship linking the transmitted field to the incident field and the susceptibilities. In order to simplify the problem, we consider here the case of a reflection-less metasurface. Inserting~\eqref{eq:field_diff} and \eqref{eq:field_av} with $\Psi_u^\text{r}=0$ ($u=x,y$) into~\eqref{eq:chi_diag} and solving for the transmitted components of the fields yields
\begin{subequations}
\label{eq:Tf}
\begin{equation}
E_{x}^{\text{t}}  =  -E_{x}^{\text{i}} + \frac{8E_{x}^{\text{i}}-j4\chi_{\text{mm}}^{yy}\mu\omega H_{y}^{\text{i}}}{4+\chi_{\text{ee}}^{xx}\chi_{\text{mm}}^{yy}\epsilon\mu\omega^2},
\end{equation}
\begin{equation}
\label{eq:Tfield2}
E_{y}^{\text{t}}  =  -E_{y}^{\text{i}} + \frac{8E_{y}^{\text{i}}+j4\chi_{\text{mm}}^{xx}\mu\omega H_{x}^{\text{i}}}{4+\chi_{\text{mm}}^{xx}\chi_{\text{ee}}^{yy}\epsilon\mu\omega^2},
\end{equation}
\begin{equation}
H_{x}^{\text{t}}  =  -H_{x}^{\text{i}} + \frac{8H_{x}^{\text{i}}+j4\chi_{\text{ee}}^{yy}\epsilon\omega E_{y}^{\text{i}}}{4+\chi_{\text{mm}}^{xx}\chi_{\text{ee}}^{yy}\epsilon\mu\omega^2},
\end{equation}
\begin{equation}
H_{y}^{\text{t}}  =  -H_{y}^{\text{i}} + \frac{8H_{y}^{\text{i}}-j4\chi_{\text{ee}}^{xx}\epsilon\omega E_{x}^{\text{i}}}{4+\chi_{\text{ee}}^{xx}\chi_{\text{mm}}^{yy}\epsilon\mu\omega^2}.
\end{equation}
\end{subequations}
These relations show how each of the transmitted field components depend on their incident field counterparts and orthogonal duals, e.g. $E_x^\text{t}=E_x^\text{t}(E_x^\text{i},H_y^\text{i})$, etc. They have to be considered after the susceptibilities \eqref{eq:chi_diag} have been synthesized for given specifications to determine whether they may be realized by a passive metasurface ($|\ve{E}^\text{t}|\le|\ve{E}^\text{i}|$ and $|\ve{H}^\text{t}|\le|\ve{H}^\text{i}|$), or require active elements.

The susceptibilities in~\eqref{eq:chi_diag} represent the synthesis (inverse problem) results of the proposed method while Eqs.~\eqref{eq:Tf} express the transmitted field components in terms of these susceptibilities (direct problem). We now need to establish the relationships existing between the susceptibilities and the scattering parameters in order to enable the second step of the synthesis (see last paragraph in Sec.~\ref{sec:metasurf_synth_prob}). The forthcoming methodology for single transformation is analogous to that proposed in~\cite{holloway2005reflection,Pfeiffer2013a,PhysRevApplied.2.044011}, while the corresponding methodologies for multiple transformation, to be presented in the next subsection, are more general.

In the plane wave approximation, which is naturally valid when the source of the incident wave is far enough from the metasurface, the response of each scattering particle may be expressed in terms of its reflection and transmission coefficients~\cite{GrbicLightBending,asadchy2011simulation,asadchy2014determining}. Since according to~\eqref{eq:Tf}, the pairs ($E_x^\text{t}$, $H_y^\text{t}$) and ($E_y^\text{t}$, $H_x^\text{t}$) are proportional to their incident counterparts and orthogonal duals only, the problem splits into an $x$-polarized incident plane wave problem and a $y$-polarized incident plane wave problem, whose fields at normal incidence are respectively given by
\begin{subequations}
\label{eq:xPol}
\begin{equation}
\label{eq:xPol1}
\ve{E}^{\text{i}}=\hat{x},
\quad\quad
\ve{E}^{\text{r}}=R_x\hat{x},
\quad\quad
\ve{E}^{\text{t}}=T_x\hat{x},
\end{equation}
\begin{equation}
\label{eq:xPol2}
\ve{H}^{\text{i}}=\frac{1}{\eta}\hat{y},
\quad\quad
\ve{H}^{\text{r}}=-\frac{R_x}{\eta}\hat{y},
\quad\quad
\ve{H}^{\text{t}}=\frac{T_x}{\eta}\hat{y},
\end{equation}
\end{subequations}
and
\begin{subequations}
\label{eq:yPol}
\begin{equation}
\label{eq:yPol1}
\ve{E}^{\text{i}}=\hat{y},
\quad\quad
\ve{E}^{\text{r}}=R_y\hat{y},
\quad\quad
\ve{E}^{\text{t}}=T_y\hat{y},
\end{equation}
\begin{equation}
\label{eq:yPol2}
\ve{H}^{\text{i}}=-\frac{1}{\eta}\hat{x},
\quad\quad
\ve{H}^{\text{r}}=\frac{R_y}{\eta}\hat{x},
\quad\quad
\ve{H}^{\text{t}}=-\frac{T_y}{\eta}\hat{x},
\end{equation}
\end{subequations}
where $R_u$ and $T_u$ ($u=x,y$) represent reflection and transmission coefficients, respectively\footnote{The waves in~\eqref{eq:xPol} and~\eqref{eq:yPol} are defined as \emph{rectilinear} (i.e. they do not change direction at the metasurface) for consistency with periodic boundary conditions to be used in full-wave simulations for the second step of the synthesis (see comment at the end of the last paragraph of Sec.~\ref{sec:gen_sol}).}. Inserting \eqref{eq:xPol} and \eqref{eq:yPol} into \eqref{eq:InvProb} with the four non-zero susceptibilities given in \eqref{eq:chi_diag} leads to the transmission and reflection coefficients
\begin{subequations}
\label{eq:Xcoef}
\begin{align}
T_x&=\frac{4+\chi_{\text{ee}}^{xx}\chi_{\text{mm}}^{yy}k^2}{(2+jk\chi_{\text{ee}}^{xx})(2+jk\chi_{\text{mm}}^{yy})},\\
R_x&=\frac{2jk\left(\chi_{\text{mm}}^{yy}-\chi_{\text{ee}}^{xx}\right)}{\left(2+jk\chi_{\text{ee}}^{xx}\right)\left(2+jk\chi_{\text{mm}}^{yy}\right)}
\end{align}
\end{subequations}
and
\begin{subequations}
\label{eq:Ycoef}
\begin{align}
T_y&=\frac{4+\chi_{\text{ee}}^{yy}\chi_{\text{mm}}^{xx}k^2}{(2+jk\chi_{\text{ee}}^{yy})(2+jk\chi_{\text{mm}}^{xx})},\\
R_y&=\frac{2jk\left(\chi_{\text{mm}}^{xx}-\chi_{\text{ee}}^{yy}\right)}{\left(2+jk\chi_{\text{ee}}^{yy}\right)\left(2+jk\chi_{\text{mm}}^{xx}\right)},
\end{align}
\end{subequations}
$k=\omega\sqrt{\mu\epsilon}=\frac{2\pi}{\lambda}$. These relations may be used in the second step of the synthesis to determine the scattering parameters corresponding to the synthesized susceptibilities. Solving~\eqref{eq:Xcoef} and \eqref{eq:Ycoef} for the susceptibilities yields
\begin{subequations}
\label{eq:chi_Sparam}
\begin{align}
\chi_{\text{ee}}^{xx}&=\frac{2j\left(T_x+R_x-1\right)}{k\left(T_x+R_x+1\right)},\\
\chi_{\text{ee}}^{yy}&=\frac{2j\left(T_y+R_y-1\right)}{k\left(T_y+R_y+1\right)},\\
\chi_{\text{mm}}^{xx}&=\frac{2j\left(T_y-R_y-1\right)}{k\left(T_y-R_y+1\right)},\\
\chi_{\text{mm}}^{yy}&=\frac{2j\left(T_x-R_x-1\right)}{k\left(T_x-R_x+1\right)}.
\end{align}
\end{subequations}
In \eqref{eq:Ycoef} and~\eqref{eq:chi_Sparam}, the reflection and transmission coefficients are associated with scattering parameters $S_{ij}$ with $i,j=1,\ldots,4$ accounting for the two ports (incident and transmitted waves) and two polarization ($x$ and $y$). Specifically, assigning ports 1, 2, 3 and 4 to $x$-polarized input, $y$-polarized input, $x$-polarized output and $y$-polarized output, respectively, one has $R_x=S_{11}$, $T_x=S_{31}$, $R_y=S_{22}$ and $T_y=S_{42}$, while the other 12 scattering parameters are not required since the chosen tensors are uniaxial so that the metasurface is not gyrotropic (i.e. does not involve transformations between $x$-polarized and $y$-polarized waves).

Examples of these transformations between the synthesized susceptibilities and the scattering parameters, as given by equations~\eqref{eq:Xcoef} and~\eqref{eq:Ycoef}, are presented in Sec.~\ref{sec:Refrac} and~\ref{ex:orbital}. As will be shown in Sec.~\ref{sec:examples}, the synthesized susceptibilities generally have both real and imaginary parts. Assuming the convention $e^{j\omega t}$, the imaginary and real parts of the susceptibilities may be associated with gain, when $\operatorname{Im}(\chi)>0$, and loss, when $\operatorname{Im}(\chi)<0$, however, this is not generally true, as compensations may exist between the imaginary parts of different susceptibility terms. It is therefore generally necessary to explicitly compute the refractive index corresponding to the considered susceptibility tensors to determine whether there is really loss or gain. For example, the metasurface described by relations~\eqref{eq:chi_diag} has the property of birefringence, with refractive indices given by
\begin{subequations}
\label{eq:refractive_index}
\begin{align}
n_x &= \sqrt{(1+\chi_{\text{ee}}^{xx})(1+\chi_{\text{mm}}^{yy})},\\
n_y &= \sqrt{(1+\chi_{\text{ee}}^{yy})(1+\chi_{\text{mm}}^{xx})}.
\end{align}
\end{subequations}
An example of such compensation will be shown in Sec.~\ref{sec:rec_nonrec_pol_rot}, where the purely imaginary electric and magnetic susceptibilities in~\eqref{eq:pol_rot_imag_chi} actually compensate each other so that the resulting refractive indices~\eqref{eq:real_indices} are purely real.

In order to simplify the metasurface unit cell design procedure, which is usually performed via full-wave simulation, one may consider ``ideal'' unit cells, i.e. unit cells made of lossless dielectric substrates and PEC metallic patterns. Applying relations~\eqref{eq:chi_Sparam} to compute the susceptibilities corresponding to such ``ideal'' unit cells will necessarily yield purely real susceptibilities since Eqs.~\eqref{eq:chi_Sparam} constitute two pairs of orthogonal TEM waves. As a consequence, the exact complex susceptibilities can not be realized, which will lead to results diverging from the specified transformation.
A solution to minimize discrepancies between the specified response and the approximated response considering only real susceptibilities is to set the imaginary parts of the susceptibilities to zero while optimizing their real parts so that the response of the metasurface follows the specified response as closely as possible. This is what has been done in~\cite{Pfeiffer2013a}, where the optimization procedure consists in minimizing the cost function
\begin{equation}
\label{eq:optim}
F = \left|T_\text{spec} - T_\text{approx} \right|^2 + \left|R_\text{spec} - R_\text{approx} \right|^2,
\end{equation}
where $T_\text{spec}$ and $R_\text{spec}$ are found from~\eqref{eq:Xcoef} (or~\eqref{eq:Ycoef} for $y$-polarized waves) using exact synthesized susceptibilities. The terms $T_\text{approx}$ and $R_\text{approx}$ are obtained from the same equations but with susceptibilities that are now purely real and have to be optimized to minimize~\eqref{eq:optim}.

\subsection{Double Transformation}
\label{sec:double_transf}

We now consider the problem of double ($T=2$) transformation [two specified wave triplets: ($\ve{\Psi}_1^\text{i}$, $\ve{\Psi}_1^\text{r}$, $\ve{\Psi}_1^\text{t}$) and ($\ve{\Psi}_2^\text{i}$, $\ve{\Psi}_2^\text{r}$, $\ve{\Psi}_2^\text{t}$)] for a mono-anisotropic ($\te{\chi}_\text{em}\equiv\te{\chi}_\text{me}=0$) but not uniaxial and hence gyrotropic metasurface. Solving the $2\times{4}=8$ equations~\eqref{eq:InvProb} for the two sets of waves under these conditions with the 2 corresponding susceptibilities per equation yields the following relations for the 8 remaining susceptibilities
\begin{subequations}
\label{eq:fullC}
\begin{equation}
\chi_{\text{ee}}^{xx}=\frac{j}{\epsilon \omega}\frac{(E_{y1,\text{av}}\Delta H_{y2}-E_{y2,\text{av}}\Delta H_{y1})}{(E_{x2,\text{av}}E_{y1,\text{av}}-E_{x1,\text{av}}E_{y2,\text{av}})},
\end{equation}
\begin{equation}
\chi_{\text{ee}}^{xy}=\frac{j}{\epsilon \omega}\frac{(E_{x2,\text{av}}\Delta H_{y1}-E_{x1,\text{av}}\Delta H_{y2})}{(E_{x2,\text{av}}E_{y1,\text{av}}-E_{x1,\text{av}}E_{y2,\text{av}})},
\end{equation}
\begin{equation}
\chi_{\text{ee}}^{yx}=\frac{j}{\epsilon \omega}\frac{(E_{y2,\text{av}}\Delta H_{x1}-E_{y1,\text{av}}\Delta H_{x2})}{(E_{x2,\text{av}}E_{y1,\text{av}}-E_{x1,\text{av}}E_{y2,\text{av}})},
\end{equation}
\begin{equation}
\chi_{\text{ee}}^{yy}=\frac{j}{\epsilon \omega}\frac{(E_{x1,\text{av}}\Delta H_{x2}-E_{x2,\text{av}}\Delta H_{x1})}{(E_{x2,\text{av}}E_{y1,\text{av}}-E_{x1,\text{av}}E_{y2,\text{av}})},
\end{equation}
\begin{equation}
\chi_{\text{mm}}^{xx}=\frac{j}{\mu \omega}\frac{(H_{y2,\text{av}}\Delta E_{y1}-H_{y1,\text{av}}\Delta E_{y2})}{(H_{x2,\text{av}}H_{y1,\text{av}}-H_{x1,\text{av}}H_{y2,\text{av}})},
\end{equation}
\begin{equation}
\chi_{\text{mm}}^{xy}=\frac{j}{\mu \omega}\frac{(H_{x1,\text{av}}\Delta E_{y2}-H_{x2,\text{av}}\Delta E_{y1})}{(H_{x2,\text{av}}H_{y1,\text{av}}-H_{x1,\text{av}}H_{y2,\text{av}})},
\end{equation}
\begin{equation}
\chi_{\text{mm}}^{yx}=\frac{j}{\mu \omega}\frac{(H_{y1,\text{av}}\Delta E_{x2}-H_{y2,\text{av}}\Delta E_{x1})}{(H_{x2,\text{av}}H_{y1,\text{av}}-H_{x1,\text{av}}H_{y2,\text{av}})},
\end{equation}
\begin{equation}
\chi_{\text{mm}}^{yy}=\frac{j}{\mu \omega}\frac{(H_{x2,\text{av}}\Delta E_{x1}-H_{x1,\text{av}}\Delta E_{x2})}{(H_{x2,\text{av}}H_{y1,\text{av}}-H_{x1,\text{av}}H_{y2,\text{av}})},
\end{equation}
\end{subequations}
where the subscripts $1$ and $2$ stand for the first and the second wave triplet transformation, respectively.

Following a similar procedure as in Sec.~\ref{sec:single_transf} but assuming that the incident wave given in~\eqref{eq:xPol1} induces not only a co-polarized but also a cross-polarized reflected ($R_{xx}$ and $R_{xy}$) and transmitted waves ($T_{xx}$ and $T_{xy}$)\footnote{For instance, the second equation in~\eqref{eq:xPol1} becomes $\ve{E}^\text{r}=(R_{xx}+R_{xy})\ve{\hat{x}}$.}. And similarly for the incident wave~\eqref{eq:yPol1} with the coefficients $R_{yy}, R_{yx}, T_{yy}$ and $T_{yx}$. Finally, the double-transformation relations between the reflection coefficients and susceptibilities read
\begin{subequations}
\begin{equation}
\begin{split}
&T_{xx} = \frac{-4-2 j  \chi_{\text{ee}}^{yy} k  }{k^2(\chi_{\text{ee}}^{xx} \chi_{\text{ee}}^{yy}-\chi_{\text{ee}}^{xy}\chi_{\text{ee}}^{yx})-2 jk (\chi_{\text{ee}}^{xx}+\chi_{\text{ee}}^{yy})-4}\\
  &+\frac{k^2(\chi_{\text{mm}}^{xx}\chi_{\text{mm}}^{yy}k  -\chi_{\text{mm}}^{xy} \chi_{\text{mm}}^{yx}k -2 j \chi_{\text{mm}}^{yy})}{k^2 (\chi_{\text{mm}}^{xy}\chi_{\text{mm}}^{yx}-\chi_{\text{mm}}^{xx} \chi_{\text{mm}}^{yy})+2 j k(\chi_{\text{mm}}^{xx}+\chi_{\text{mm}}^{yy})+4},
\end{split}
\end{equation}
\begin{equation}
\begin{split}
&T_{xy} =  \frac{ 2 j k\chi_{\text{ee}}^{xy}}{k^2 (\chi_{\text{ee}}^{xx}
   \chi_{\text{ee}}^{yy}-\chi_{\text{ee}}^{xy} \chi_{\text{ee}}^{yx})-2 jk
   (\chi_{\text{ee}}^{xx}+\chi_{\text{ee}}^{yy})-4}\\
   &+\frac{ 2 j k\chi_{\text{mm}}^{yx} }{k ^2 (\chi_{\text{mm}}^{xy} \chi_{\text{mm}}^{yx}-\chi_{\text{mm}}^{xx} \chi_{\text{mm}}^{yy})+2 j k(\chi_{\text{mm}}^{xx}+\chi_{\text{mm}}^{yy})+4},
\end{split}
\end{equation}
\begin{equation}
\begin{split}
& T_{yx} = \frac{2 j k \chi_{\text{ee}}^{yx} }{k^2 (\chi_{\text{ee}}^{xx}
   \chi_{\text{ee}}^{yy}-\chi_{\text{ee}}^{xy} \chi_{\text{ee}}^{yx})-2 jk
   (\chi_{\text{ee}}^{xx}+\chi_{\text{ee}}^{yy})-4}\\
   &+\frac{2 j k \chi_{\text{mm}}^{xy} }{k^2 (\chi_{\text{mm}}^{xy} \chi_{\text{mm}}^{yx}-\chi_{\text{mm}}^{xx} \chi_{\text{mm}}^{yy})+2
   j k  (\chi_{\text{mm}}^{xx}+\chi_{\text{mm}}^{yy})+4 },
\end{split}
\end{equation}
\begin{equation}
\begin{split}
&T_{yy} = \frac{-4-2 j \chi_{\text{ee}}^{xx} k }{k^2(\chi_{\text{ee}}^{xx} \chi_{\text{ee}}^{yy}-\chi_{\text{ee}}^{xy} \chi_{\text{ee}}^{yx})-2 jk
   (\chi_{\text{ee}}^{xx}+\chi_{\text{ee}}^{yy})-4}\\
   &+\frac{k^2(\chi_{\text{mm}}^{xx}
   \chi_{\text{mm}}^{yy}k -\chi_{\text{mm}}^{xy}
   \chi_{\text{mm}}^{yx}k -2 j \chi_{\text{mm}}^{xx} )}{k^2 (\chi_{\text{mm}}^{xy}
   \chi_{\text{mm}}^{yx}-\chi_{\text{mm}}^{xx} \chi_{\text{mm}}^{yy})+2 jk
   (\chi_{\text{mm}}^{xx}+\chi_{\text{mm}}^{yy})+4},
   \end{split}
\end{equation}
\begin{floatEq}
\begin{equation}
R_{xx} = \frac{2k \left((2 +j \chi_{\text{mm}}^{xx}k )
   \left((k(\chi_{\text{ee}}^{xy} \chi_{\text{ee}}^{yx}-\chi_{\text{ee}}^{xx} \chi_{\text{ee}}^{yy})+2j\chi_{\text{ee}}^{xx})+\chi_{\text{mm}}^{yy}(\chi_{\text{ee}}^{yy}k -2 j)\right)-\chi_{\text{mm}}^{xy} \chi_{\text{mm}}^{yx}k(2+j\chi_{\text{ee}}^{yy} k
   )\right)}{(k^2 (\chi_{\text{ee}}^{xx} \chi_{\text{ee}}^{yy}-\chi_{\text{ee}}^{xy} \chi_{\text{ee}}^{yx})-2jk(\chi_{\text{ee}}^{xx}+\chi_{\text{ee}}^{yy})-4) \left(k^2(\chi_{\text{mm}}^{xy} \chi_{\text{mm}}^{yx}-\chi_{\text{mm}}^{xx} \chi_{\text{mm}}^{yy})+2jk(\chi_{\text{mm}}^{xx}+\chi_{\text{mm}}^{yy})+4\right)},
\end{equation}
\begin{equation}
R_{yy} = \frac{2k\left((2 +j\chi_{\text{mm}}^{yy}k)\left((k   (\chi_{\text{ee}}^{xy}\chi_{\text{ee}}^{yx}-\chi_{\text{ee}}^{xx} \chi_{\text{ee}}^{yy})+2 j\chi_{\text{ee}}^{yy})+\chi_{\text{mm}}^{xx}(\chi_{\text{ee}}^{xx}k -2 j)\right)-\chi_{\text{mm}}^{xy} \chi_{\text{mm}}^{yx} k(2+j\chi_{\text{ee}}^{xx}k)\right)}{(k^2(\chi_{\text{ee}}^{xx} \chi_{\text{ee}}^{yy}-\chi_{\text{ee}}^{xy} \chi_{\text{ee}}^{yx})-2 jk
   (\chi_{\text{ee}}^{xx}+\chi_{\text{ee}}^{yy})-4)\left(k^2(\chi_{\text{mm}}^{xy} \chi_{\text{mm}}^{yx}-\chi_{\text{mm}}^{xx}\chi_{\text{mm}}^{yy})+2jk(\chi_{\text{mm}}^{xx}+\chi_{\text{mm}}^{yy})+4\right)},
\end{equation}
\end{floatEq}
\begin{equation}
\begin{split}
&R_{xy} = \frac{2 j k\chi_{\text{ee}}^{xy} }{k^2(\chi_{\text{ee}}^{xx} \chi_{\text{ee}}^{yy}-\chi_{\text{ee}}^{xy}
   \chi_{\text{ee}}^{yx})-2 jk
   (\chi_{\text{ee}}^{xx}+\chi_{\text{ee}}^{yy})-4}\\
   &-\frac{2 j k\chi_{\text{mm}}^{yx}}{k^2 (\chi_{\text{mm}}^{xy} \chi_{\text{mm}}^{yx}-\chi_{\text{mm}}^{xx} \chi_{\text{mm}}^{yy})+2
   jk(\chi_{\text{mm}}^{xx}+\chi_{\text{mm}}^{yy})+4},
   \end{split}
\end{equation}
\begin{equation}
\begin{split}
&R_{yx} = \frac{2 j k\chi_{\text{ee}}^{yx}}{k^2(\chi_{\text{ee}}^{xx} \chi_{\text{ee}}^{yy}-\chi_{\text{ee}}^{xy}
   \chi_{\text{ee}}^{yx})-2 j
   (\chi_{\text{ee}}^{xx}+\chi_{\text{ee}}^{yy})-4}\\
   &-\frac{2 jk\chi_{\text{mm}}^{xy}}{k^2 (\chi_{\text{mm}}^{xy} \chi_{\text{mm}}^{yx}-\chi_{\text{mm}}^{xx} \chi_{\text{mm}}^{yy})+2
   jk(\chi_{\text{mm}}^{xx}+\chi_{\text{mm}}^{yy})+4}.
   \end{split}
\end{equation}
\end{subequations}
In these relations, assigning ports 1, 2, 3 and 4 to $x$-polarized input, $y$-polarized input, $x$-polarized output and $y$-polarized output, respectively, one has $T_{xx}=S_{31}$, $T_{yx}=S_{41}$, $T_{yy}=S_{42}$ and $T_{xy}=S_{32}$.

Note that the optimization procedure described in Sec.~\ref{sec:single_transf} where~\eqref{eq:optim} is minimized has to be modified so as to include all the transmission and reflection coefficients relating to each susceptibility as follows
\begin{equation}
 F = \sum_{u,v \in \{x,y\}} |T^{uv}_\text{spec} - T^{uv}_\text{approx}|^2+|R^{uv}_\text{spec} - R^{uv}_\text{approx}|^2
\end{equation}

\subsection{Solutions for Normal Polarization Densities}
\label{sec:normal_synthesis}

Solving the general bi-anisotropic system of equations~\eqref{eq:BC} in the case of full electric and magnetic susceptibility tensors (i.e. nonzero $\te{\chi}_{\text{ee}},\te{\chi}_{\text{mm}},\te{\chi}_{\text{me}}$ and $\te{\chi}_{\text{em}}$) leads to a total number of 36~unknowns for only 6~equations. This corresponds to a heavily under-determined system for single-wave transformation. This means that the system has effectively the capability to handle up to 6~wave triplets, or transformations, in the most general case or more depending on the choice of susceptibilities and the specified transformations (e.g. no rotation of polarization).
For single-wave transformation, only 6 (or even less) independent susceptibilities may be set to nonzero values, since this would lead to an exactly determined system of equal number of equations and unknowns.

Let us consider the case where $\te{\chi}_{\text{me}}=\te{\chi}_{\text{em}}=0$ and $\te{\chi}_{\text{ee}}$ and $\te{\chi}_{\text{mm}}$ are diagonal. This corresponds to the polarization densities
\begin{subequations}
\label{eq:Pz}
\begin{align}
\ve{P}&=\epsilon
\begin{pmatrix}
\chi_{\text{ee}}^{xx} & 0 & 0 \\
0 & \chi_{\text{ee}}^{yy} & 0 \\
0 & 0 & \chi_{\text{ee}}^{zz}
\end{pmatrix}
\begin{pmatrix}
E_{x,\text{av}}\\
E_{y,\text{av}}\\
E_{z,\text{av}}
\end{pmatrix},\\
\label{eq:Mz}
\ve{M}&=
\begin{pmatrix}
\chi_{\text{mm}}^{xx} & 0 & 0 \\
0 & \chi_{\text{mm}}^{yy} & 0 \\
0 & 0 & \chi_{\text{mm}}^{zz}
\end{pmatrix}
\begin{pmatrix}
H_{x,\text{av}}\\
H_{y,\text{av}}\\
H_{z,\text{av}}
\end{pmatrix}.
\end{align}
\end{subequations}
Inserting~\eqref{eq:Pz} into~\eqref{eq:BC} and solving for the susceptibilities that do not fall under derivatives yields the following system of equations
\begin{subequations}
\label{eq:Coupled_sys}
\begin{align}
\chi_{\text{ee}}^{xx}&=\frac{1}{j\omega\epsilon E_{x,\text{av}}}\left[-\Delta H_{y}-\frac{\partial}{\partial y}\left(\chi_{\text{mm}}^{zz}H_{z,\text{av}} \right)\right],\label{eq:Coupled_sys_1}\\
\chi_{\text{ee}}^{yy}&=\frac{1}{j\omega\epsilon E_{y,\text{av}}}\left[\Delta H_{x}+\frac{\partial}{\partial x}\left(\chi_{\text{mm}}^{zz}H_{z,\text{av}} \right)\right],\label{eq:Coupled_sys_2}\\
\chi_{\text{ee}}^{zz}&=-\frac{1}{E_{z,\text{av}}}\left[\Delta E_{z}+\frac{\partial}{\partial x}\left(\chi_{\text{ee}}^{xx}E_{x,\text{av}} \right)+\frac{\partial}{\partial y}\left(\chi_{\text{ee}}^{yy}E_{y,\text{av}} \right)\right],\label{eq:Coupled_sys_3}\\
\chi_{\text{mm}}^{xx}&=\frac{1}{j\omega\mu H_{x,\text{av}}}\left[\Delta E_{y}+\frac{\partial}{\partial y}\left(\chi_{\text{ee}}^{zz}E_{z,\text{av}} \right)\right],\label{eq:Coupled_sys_4}\\
\chi_{\text{mm}}^{yy}&=\frac{1}{j\omega\mu H_{y,\text{av}}}\left[-\Delta E_{x}-\frac{\partial}{\partial x}\left(\chi_{\text{ee}}^{zz}E_{z,\text{av}} \right)\right],\label{eq:Coupled_sys_5}\\
\chi_{\text{mm}}^{zz}&=-\frac{1}{H_{z,\text{av}}}\left[\Delta H_{z}+\frac{\partial}{\partial x}\left(\chi_{\text{mm}}^{xx}H_{x,\text{av}} \right)+\frac{\partial}{\partial y}\left(\chi_{\text{mm}}^{yy}H_{y,\text{av}} \right)\right].\label{eq:Coupled_sys_6}
\end{align}
\end{subequations}
This set of coupled inhomogenous partial differential equations can be simplified by substituting~\eqref{eq:Coupled_sys_1} and~\eqref{eq:Coupled_sys_2} into~\eqref{eq:Coupled_sys_3} and~\eqref{eq:Coupled_sys_4} and~\eqref{eq:Coupled_sys_5} into~\eqref{eq:Coupled_sys_6}. The normal component of the electric and magnetic susceptibilities are subsequently given by
\begin{subequations}
\label{eq:Coupled_sys_reduced}
\begin{align}
\chi_{\text{ee}}^{zz}&=-\frac{1}{E_{z,\text{av}}}\left[\Delta E_{z}-
\frac{1}{j\omega\epsilon}\left( \frac{\partial\Delta H_y}{\partial x} - \frac{\partial\Delta H_x}{\partial y}\right)\right],\\
\chi_{\text{mm}}^{zz}&=-\frac{1}{H_{z,\text{av}}}\left[\Delta H_{z}+
\frac{1}{j\omega\mu}\left( \frac{\partial\Delta E_y}{\partial x} -\frac{\partial\Delta E_x}{\partial y} \right)\right],
\end{align}
\end{subequations}
whose right-hand side expressions in square brackets vanish as they correspond to the left-hand sides of Maxwell equations written as $\ve{E}-\frac{1}{j\omega\epsilon}\nabla\times\ve{H}=0$ and $\ve{H}+\frac{1}{j\omega\mu}\nabla\times\ve{E}=0$, respectively, on each side of the metasurface, i.e. \mbox{$\chi_{\text{ee}}^{zz}=\chi_{\text{mm}}^{zz}=0$}. In this case, the system~\eqref{eq:Coupled_sys} has reduced to the system~\eqref{eq:chi_diag}.\\
This result might come as a surprise. However, it could have been expected by the following consideration. In the case of a uniaxially anisotropic surface, specification of the tangential fields following boundary conditions fully determines the normal fields. Therefore, if one were allowed to chose $\chi_{\text{ee}}^{zz}$ or $\chi_{\text{mm}}^{zz}$, one would alter the normal fields, and hence violate Maxwell equations. Thus, these susceptibilities must be zero.

\begin{figure}[h!]
\centering
\includegraphics[width=0.75\linewidth]{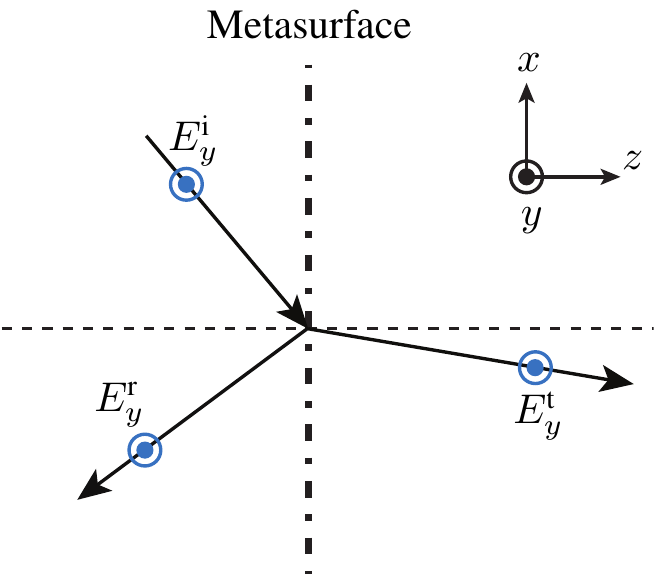}
\caption{Generalized refraction.}
\label{eq:refrac_Pz}
\end{figure}

Consider now the case of single-transformation generalized refraction, illustrated in Fig.~\ref{eq:refrac_Pz}, where the metasurface arbitrarily reflects and ``refracts'' an arbitrary incident wave. Assume that all the waves are polarized along $y$ (perpendicular polarization), so that $E_x=E_z=0$ and $H_y=0$, which corresponds to a 2D problem with $\frac{\partial}{\partial y}=0$.

In order to obtain closed-form expressions for the susceptibilities, a determined system of equations has to be chosen. To reduce the number of unknowns, consider $\te{\chi}_{\text{me}}=\te{\chi}_{\text{em}}=0$. Keeping only the relevant susceptibility components of $\ve{P}$ and $\ve{M}$ yields then
\begin{subequations}
\label{eq:Pz2}
\begin{align}
\ve{P}&=\epsilon
\begin{pmatrix}
0 & 0 & 0 \\
0 & \chi_{\text{ee}}^{yy} & 0 \\
0 & 0 & 0
\end{pmatrix}
\begin{pmatrix}
0\\
E_{y,\text{av}}\\
0
\end{pmatrix},\\
\label{eq:Mz2}
\ve{M}&=
\begin{pmatrix}
\chi_{\text{mm}}^{xx} & 0 & \chi_{\text{mm}}^{xz} \\
0 & 0 & 0 \\
\chi_{\text{mm}}^{zx} & 0 & \chi_{\text{mm}}^{zz}
\end{pmatrix}
\begin{pmatrix}
H_{x,\text{av}}\\
0\\
H_{z,\text{av}}
\end{pmatrix}.
\end{align}
\end{subequations}
It would be erroneous to think that the corresponding system~\eqref{eq:BC} is still composed of 6~independent equations. The system~\eqref{eq:BC} in this case reduces indeed to the only 3~following equations
\begin{subequations}
\label{eq:reduced_BC}
\begin{align}
\Delta H_x={}&j\omega\epsilon\chi_{\text{ee}}^{yy}E_{y,\text{av}}-\frac{\partial}{\partial x}\left(\chi_{\text{mm}}^{zx}H_{x,\text{av}}+\chi_{\text{mm}}^{zz}H_{z,\text{av}} \right),\\
\Delta E_y={}&j\omega\mu\left(\chi_{\text{mm}}^{xx}H_{x,\text{av}}+\chi_{\text{mm}}^{xz}H_{z,\text{av}} \right),\label{eq:reduced_BC_2}\\
\begin{split}
\Delta H_z={}&-\chi_{\text{mm}}^{zx}H_{x,\text{av}}-\chi_{\text{mm}}^{zz}H_{z,\text{av}}\\
            &-\frac{\partial}{\partial x}\left(\chi_{\text{mm}}^{xx}H_{x,\text{av}}+\chi_{\text{mm}}^{xz}H_{z,\text{av}} \right),
\end{split}
\end{align}
\end{subequations}
which constitute a system of 3~equations for the 5~unknown susceptibilities in~\eqref{eq:Pz2}. Since this is an under-determined system, it cannot be solved without further specifications. Consider therefore the four following restricted cases:
\begin{enumerate}
\item Case I: $\chi_{\text{ee}}^{yy}, \chi_{\text{mm}}^{xx}$ and $\chi_{\text{mm}}^{zz}$.
\item Case II: $\chi_{\text{ee}}^{yy}, \chi_{\text{mm}}^{xx}$ and $\chi_{\text{mm}}^{zx}=\chi_{\text{mm}}^{xz}$ (reciprocity condition~\eqref{eq:reciprocity}).
\item Case III: $\chi_{\text{ee}}^{yy}, \chi_{\text{mm}}^{zz}$ and $\chi_{\text{mm}}^{zx}=\chi_{\text{mm}}^{xz}$ (reciprocity condition~\eqref{eq:reciprocity}).
\item Case IV: $\chi_{\text{ee}}^{yy}$ and $\chi_{\text{mm}}^{zz}$.
\end{enumerate}
Case I is a purely diagonal case. Cases II and III reduce the number of independent unknowns by enforcing reciprocity ~\eqref{eq:reciprocity}. In case IV, the presence of the only two maintained susceptibilities eliminates~\eqref{eq:reduced_BC_2}. Inserting the electric and magnetic polarization densities corresponding to these 4~cases into~\eqref{eq:reduced_BC} yields the following closed-form relations.
\subsubsection{Case I}
\begin{subequations}
\label{eq:Case_I}
\begin{align}
\chi_{\text{mm}}^{xx}&=\frac{\Delta E_y}{j\omega\mu H_{x,\text{av}}},\label{eq:Case_I_1}\\
\chi_{\text{mm}}^{zz}&=-\frac{1}{H_{z,\text{av}}}\left[\Delta H_z+\frac{\partial}{\partial x}\left( \frac{\Delta E_y}{j\omega\mu} \right) \right],\label{eq:Case_I_2}\\
\chi_{\text{ee}}^{yy}&=\frac{1}{j\omega\epsilon E_{y,\text{av}}}\left\{\Delta H_x- \frac{\partial}{\partial x}\left[ \Delta H_z + \frac{\partial}{\partial x}\left( \frac{\Delta E_y}{j\omega\mu} \right)\right] \right\}.\label{eq:Case_I_3}
\end{align}
\end{subequations}
The expression inside the bracket of~\eqref{eq:Case_I_2} corresponds to the Maxwell-Faraday equation. It follows that~\eqref{eq:Case_I_2} simplifies to $\chi_{\text{mm}}^{zz}=0$. The same consideration applies to~\eqref{eq:Case_I_3}, leaving only $\Delta H_x$ in the curl bracket. Comparing the remaining equations with~\eqref{eq:chi_diag}, one finds that~\eqref{eq:Case_I_1} is equal to~\eqref{eq:chi_diag_Mxx}, while~\eqref{eq:Case_I_3} reduces to~\eqref{eq:chi_diag_Eyy}. Thus, the reason why $\chi_{\text{mm}}^{zz}=0$ is the same as that given in the previous problem.
\subsubsection{Case II}
\begin{subequations}
\label{eq:Case_II}
\begin{align}
\chi_{\text{mm}}^{xx}&=\frac{\Delta E_y}{j\omega\mu H_{x,\text{av}}}+\frac{H_{z,\text{av}}}{H_{x,\text{av}}^2}\left[\Delta H_z +\frac{\partial}{\partial x}\left( \frac{\Delta E_y}{j\omega\mu} \right) \right],\label{eq:Case_II_1}\\
\chi_{\text{mm}}^{zx}&=\chi_{\text{mm}}^{xz}=-\frac{1}{H_{x,\text{av}}}\left[\Delta H_z+\frac{\partial}{\partial x}\left( \frac{\Delta E_y}{j\omega\mu} \right)\right],\label{eq:Case_II_2}\\
\chi_{\text{ee}}^{yy}&=\frac{1}{j\omega\epsilon E_{y,\text{av}}}\left\{\Delta H_x- \frac{\partial}{\partial x}\left[ \Delta H_z + \frac{\partial}{\partial x}\left( \frac{\Delta E_y}{j\omega\mu} \right)\right] \right\}.\label{eq:Case_II_3}
\end{align}
\end{subequations}
For the same reasons as in the previous case, $\chi_{\text{mm}}^{zx}=\chi_{\text{mm}}^{xz}=0$, and the resulting system is identical to that of Case I.
\subsubsection{Case III}
\begin{subequations}
\label{eq:Case_III}
\begin{align}
\chi_{\text{mm}}^{zx}&=\chi_{\text{mm}}^{xz}=\frac{\Delta E_y}{j\omega\mu H_{z,\text{av}}},\\
\chi_{\text{mm}}^{zz}&=\frac{-\Delta E_y}{j\omega\mu}\frac{H_{x,\text{av}}}{H_{z,\text{av}}^2},\\
\chi_{\text{ee}}^{yy}&=\frac{\Delta H_x}{j\omega\epsilon E_{y,\text{av}}}.\label{eq:Case_III_3}
\end{align}
\end{subequations}
These relations have been simplified using Maxwell-Faraday equation. It appears that the electric susceptibility~\eqref{eq:Case_III_3} is identical to that in~\eqref{eq:chi_diag_Eyy}. In this case, $\chi_{\text{mm}}^{zx}\neq 0$ and $\chi_{\text{mm}}^{xz}\neq 0$ since the $x$-component of the magnetic field has not been specified, and is therefore specified through the normal terms.
\subsubsection{Case IV}
\begin{subequations}
\label{eq:Case_IV}
\begin{align}
\chi_{\text{mm}}^{zz}&=\frac{-\Delta H_z}{H_{z,\text{av}}},\label{eq:Case_IV_1}\\
\chi_{\text{ee}}^{yy}&=\frac{1}{j\omega\epsilon E_{y,\text{av}}}\left[\Delta H_x - \frac{\partial}{\partial x}\left( \Delta H_z \right)\right],\label{eq:Case_IV_2}\\
&=\frac{\Delta H_x}{j\omega\epsilon E_{y,\text{av}}}+\frac{\Delta E_y}{E_{y,\text{av}}}.\label{eq:Case_IV_3}
\end{align}
\end{subequations}
Here the electric susceptibility given in~\eqref{eq:Case_IV_2} has been further simplified to~\eqref{eq:Case_IV_3} using Maxwell-Amp\`ere equation. This equation is identical to~\eqref{eq:chi_diag_Eyy} plus a term that is similar to~\eqref{eq:Case_IV_1}.

\section{Illustrative Examples}
\label{sec:examples}

The diversity of possible wave transformations using the proposed metasurface synthesis method is virtually infinite. This section presents a few examples of practical interest, in order of increasing complexity. In all cases, the reflection is enforced to be zero, i.e. $\Psi_u^\text{r}=0$ for $\ve{\Psi}=\ve{E},\ve{H}$ and $u=x,y,z$. Introduction of reflection is straightforward, consisting simply in adding specified $\Psi_u^\text{r}$'s in~\eqref{eq:field_diff} and~\eqref{eq:field_av}, and typically reduces the design constraints of the metasurface. The metasurface is also infinite in all the examples. Restricting the dimension of the metasurface to $L_x\times L_y$ is straightforwardly accomplished by specifying all field quantities over this area and setting them to zero elsewhere in~\eqref{eq:chi_diag} or~\eqref{eq:fullC}.

\subsection{Generalized Refraction}
\label{sec:Refrac}
\noindent \emph{Problem:} Synthesize a metasurface transforming an obliquely incident plane wave forming a $\pi/8$ angle with respect to $z$ in the $x-z$ plane into a transmitted plane wave with a $\pi/3$ ``refraction'' angle (TM refraction).

\begin{figure}[h!]
\subfloat[]{\includegraphics[width=0.45\linewidth]{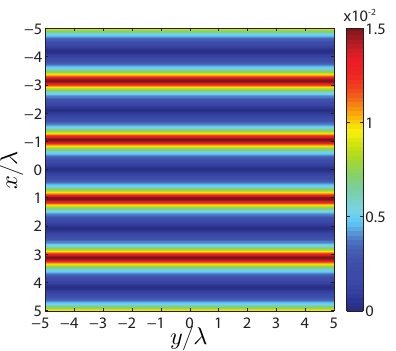}}
\subfloat[]{\includegraphics[width=0.45\linewidth]{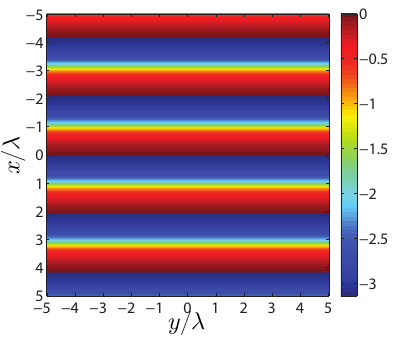}}\\
\subfloat[]{\includegraphics[width=0.45\linewidth]{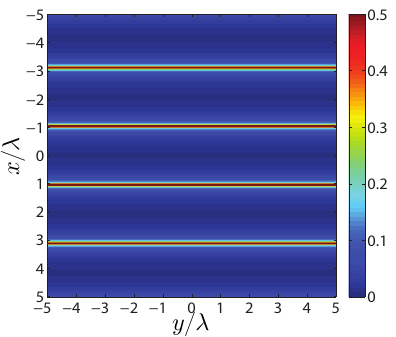}}
\subfloat[]{\includegraphics[width=0.45\linewidth]{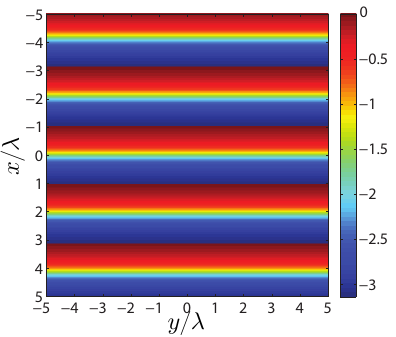}}
\caption{Magnitude (a) and phase (b) of $\chi_{\text{ee}}^{xx}(x,y)$, magnitude (c) and phase (d) of $\chi_{\text{mm}}^{yy}(x,y)$ for a metasurface refracting a plane wave incident with an angle of $\pi/8$ with respect to $z$ in the $x-z$ plane into a transmitted plane wave forming a $\pi/3$ angle.}
\label{eq:susc_refr}
\end{figure}

\begin{figure}[h!]
\subfloat[]{\includegraphics[width=0.45\linewidth]{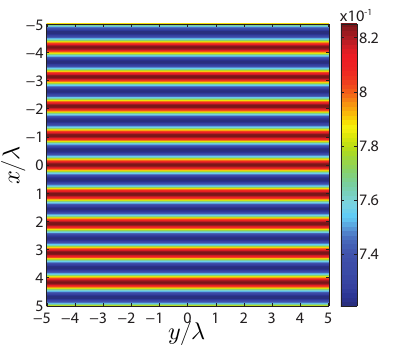}}
\subfloat[]{\includegraphics[width=0.448\linewidth]{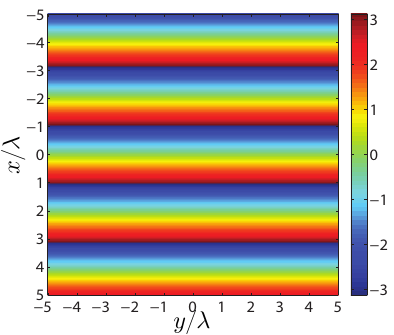}\label{fig:phaseTx_refrac}}\\
\subfloat[]{\includegraphics[width=0.45\linewidth]{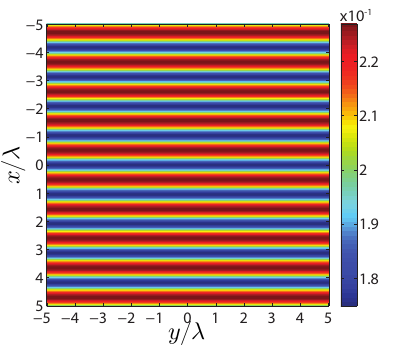}}
\subfloat[]{\includegraphics[width=0.45\linewidth]{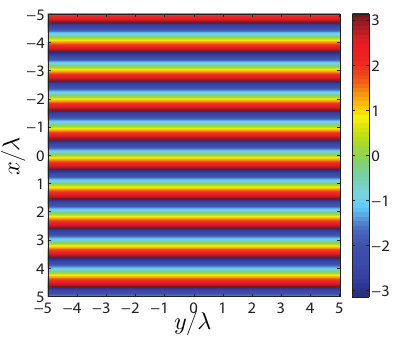}}
\caption{Magnitude (a) and phase (b) of $T_x(x,y)$, magnitude (c) and phase (d) of $R_x(x,y)$ for the problem of Fig.~\ref{eq:susc_refr}.}
\label{fig:TRcoef_refr}
\end{figure}

\noindent \emph{Synthesis:} A unit-amplitude incident plane wave (waveform $e^{j\left[\omega t-k\left(\sin(\pi/8)x + \cos(\pi/8)z\right)\right]}$) impinging on the metasurface plane ($z=0$) with a $\pi/8$ angle in the $x-z$ plane reads

\begin{subequations}
\begin{align}
\ve{E}^{\text{i}}(x,y)&=\ve{\hat{x}}\left[\cos(\pi/8)-\ve{\hat{z}}\sin(\pi/8)\right]e^{-jk\sin(\pi/8)x},\\
\ve{H}^{\text{i}}(x,y)&=\ve{\hat{y}}\frac{1}{\eta}e^{-jk\sin(\pi/8)x},
\end{align}
\end{subequations}

where $\eta=\sqrt{\mu/\epsilon}$ is the intrinsic impedance of the surrounding medium. The specified refracted wave (waveform $e^{j[\omega t-k\sqrt{2}(x+z)/2]}$), assuming zero reflection ($\ve{E}^{\text{r}}\equiv\ve{H}^{\text{r}}\equiv{0}$), is at $z=0$
\begin{subequations}
\begin{align}
\ve{E}^{\text{t}}(x,y)&=(\ve{\hat{x}}\frac{1}{2}-\ve{\hat{z}}\frac{\sqrt{3}}{2})e^{-jk\frac{\sqrt{2}}{2}x},\\
\ve{H}^{\text{t}}(x,y)&=\ve{\hat{y}}\frac{1}{\eta}e^{-jk\frac{\sqrt{2}}{2}x}.
\end{align}
\end{subequations}
The corresponding difference and average fields in~\eqref{eq:field_diff} and~\eqref{eq:field_av} are, for the components in the plane of the metasurface, given by
\begin{subequations}
\label{eq:refr_susc}
\begin{align}
\Delta E_x&=\frac{1}{2}e^{-jk\frac{\sqrt{2}}{2}x}-\cos(\pi/8)e^{-jk\sin(\pi/8)x},\\
\Delta H_y&=\frac{1}{\eta}\left(e^{-jk\frac{\sqrt{2}}{2}x}-e^{-jk\sin(\pi/8)x}\right),\\
E_{\text{av},x}&=\frac{1}{2}\left(\frac{1}{2}e^{-jk\frac{\sqrt{2}}{2}x}+\cos(\pi/8)e^{-jk\sin(\pi/8)x}\right),\\
H_{\text{av},y}&=\frac{1}{2\eta}\left(e^{-jk\frac{\sqrt{2}}{2}x}+e^{-jk\sin(\pi/8)x}\right),
\end{align}
\end{subequations}
The (non-zero) metasurface susceptibilities are then obtained by substituting~\eqref{eq:refr_susc} into~\eqref{eq:chi_diag}. They are naturally obtained in closed-form given the closed-forms~\eqref{eq:refr_susc}, but are not written explicitly here, for conciseness. Instead, the susceptibilities $\chi_\text{ee}^{xx}(x,y)$ and $\chi_\text{mm}^{yy}(x,y)$ are plotted in Fig.~\ref{eq:susc_refr} over a metasurface area of $10\lambda\times{10\lambda}$ for illustration. Note that the susceptibilities for this problem only depend on $x$, since no wave transformation is prescribed in the $y$ direction. Moreover, only two of the four susceptibilities ($\chi_\text{ee}^{xx}$ and $\chi_\text{mm}^{yy}$) are required (the other two being undefined) since no electric fields exist along \emph{y} and no magnetic fields exist along \emph{x}. Also note that the $x$-periodicity of the susceptibility is larger than $\lambda$, suggesting that this metasurface should be easily implementable with simple scattering particles.
The corresponding transmission and reflection coefficients computed using~\eqref{eq:Xcoef} are shown in Fig.~\ref{fig:TRcoef_refr}. As expected from the above discussion, the coefficients $R_y$ and $T_y$ do not exist since the field polarizations do not rotate across the metasurface. As may be seen, the transmission coefficient oscillates around $~77\%$ while the reflection coefficient oscillates around $~20\%$. The fact that the observed reflection coefficient is non-zero may a priori appear contradictory given the prescription of zero reflection. However, remember that the scattering coefficients are computed based on the assumption of rectilinear propagation, which obviously does not correspond to the present example. The actual reflection produced by the susceptibilities plotted in Fig.~\eqref{eq:susc_refr} is \emph{rigorously zero}, and the non-zero reflection parameter in Fig.~\ref{fig:TRcoef_refr} is an artifact of the mapping between the rectilinear scattering parameters and the physical problem. However, as pointed out in the last paragraph of Sec.~\ref{sec:gen_sol}, these scattering parameters can be directly used for synthesis: full-wave (using periodic boundary conditions) designing the scattering particles so that they produce the same scattering parameters as those obtained using~\eqref{eq:Xcoef} will automatically provide the desired physical solution in real (non-rectilinear) conditions.

Using the optimization procedure proposed at the end of Sec.\ref{sec:single_transf}, where the susceptibilities are forced to real values, approximate reflection and transmission coefficients ($R_\text{approx}$ and $T_\text{approx}$, respectively) can be found by minimizing~\eqref{eq:optim}. The magnitude of the new transmission coefficient (not shown here) is now very close to full and quasi-uniform transmission ($T\approx100\%,\forall x,y$) while the phase profile of the transmission coefficient has remained unaltered compared to that computed from the exact susceptibilities presented in Fig.~\ref{fig:phaseTx_refrac}. Since the transmission phase has not been changed from the exact one, the approximate metasurface performs a transformation that essentially follows the specification, except for undesired diffraction orders due to the nullification of the imaginary parts of the susceptibilities~\cite{Pfeiffer2013a}.

\subsection{Reciprocal and Non-reciprocal Polarization Rotation}
\label{sec:rec_nonrec_pol_rot}

\begin{figure}[h!]
\includegraphics[width=\linewidth]{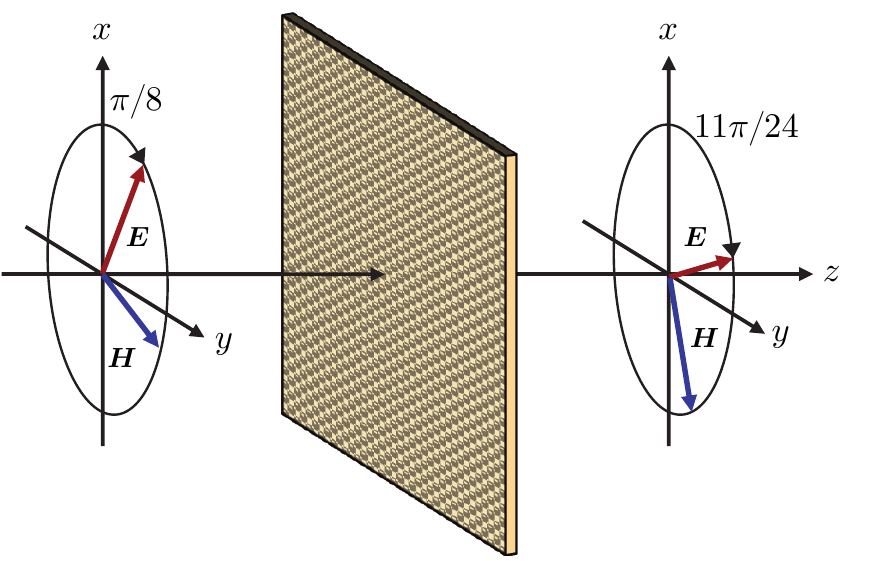}
\caption{Illustration of the problem of Sec.~\eqref{sec:rec_nonrec_pol_rot}: The  polarization of a normally incident plane wave linearly polarized at a $\pi/8$ angle with respect to the $x$-axis is rotated by $\pi/3$. In the reciprocal case, the fields retrieve their initial polarization upon propagation along the negative $z$-direction, while in the nonreciprocal case the fields experience a round-trip rotation of $2\pi/3$.}
\label{fig:Faraday_rot1}
\end{figure}

\noindent \emph{Problem:} Synthesize a metasurface transforming a normally incident linearly polarized plane wave with the electric field making an angle of $\pi/8$ with respect to the $x$-axis into a normally transmitted plane wave whose fields are rotated by $\pi/3$ (total angle of $11\pi/24$ with respect to the $x$-axis), as illustrated in Fig.~\ref{fig:Faraday_rot1}. Consider both the reciprocal and non-reciprocal cases.

\noindent \emph{Synthesis:} The prescribed incident and transmitted fields at $z=0$ read
\begin{subequations}
\begin{align}
\ve{E}^{\text{i}}(x,y)&=\ve{\hat{x}}\cos(\pi/8)+\ve{\hat{y}}\sin(\pi/8),\\
\ve{H}^{\text{i}}(x,y)&=\frac{1}{\eta}\left[-\ve{\hat{x}}\sin(\pi/8)+\ve{\hat{y}}\cos(\pi/8)\right],
\end{align}
\end{subequations}
and
\begin{subequations}
\begin{align}
\ve{E}^{\text{t}}(x,y)&=\ve{\hat{x}}\cos(11\pi/24)+\ve{\hat{y}}\sin(11\pi/24),\\
\ve{H}^{\text{i}}(x,y)&=\frac{1}{\eta}\left[-\ve{\hat{x}}\sin(11\pi/24)+\ve{\hat{y}}\cos(11\pi/24)\right],
\end{align}
\end{subequations}
respectively. From this point, the difference and average fields are straightforwardly found as in~\eqref{eq:refr_susc}. Note that in this problem the fields are independent on $x$ and $y$ since the transformation only affects the direction of the field polarizations and does not alter the direction of propagation. Therefore, the synthesized susceptibilities are constant.

In the reciprocal scenario, the metasurface is a chiral surface. While the polarization of the wave sent along the positive $z$-direction is rotated by an angle $\pi/3$, as it is sent back along the negative $z$-direction from the its transmitted angle, it returns to its original direction (here $\pi/8$ with respect to $x$). Therefore, the (non-zero) metasurface susceptibilities are again obtained by substitution of the difference and average fields into~\eqref{eq:chi_diag}, which yields
\begin{subequations}
\label{eq:chi_reciprocal_rot}
\begin{align}
\chi_{\text{ee}}^{xx}&=\chi_{\text{mm}}^{yy}=-0.0239j,\label{eq:chi_reciprocal_rot_a}\\
\chi_{\text{ee}}^{yy}&=\chi_{\text{mm}}^{xx}=0.0141j.\label{eq:chi_reciprocal_rot_b}
\end{align}
\end{subequations}
Because the diagonal components of the susceptibility tensors are used in this example, the synthesized metasurface is not able to directly rotate the polarization of the incident wave. Rather, it reduces the $x$-component of the electric field and amplifies its $y$-component. This is manifested first by purely imaginary susceptibilities in~\eqref{eq:chi_reciprocal_rot}, and secondly by opposite signs in~\eqref{eq:chi_reciprocal_rot_a} (negative sign related to absorption) and~\eqref{eq:chi_reciprocal_rot_b} (positive sign related to gain). The refractive indices corresponding to this metasurface are obtained by~\eqref{eq:refractive_index} as
\begin{subequations}
\label{eq:chi_reciprocal_rot_refrac}
\begin{align}
n_x &= 1-0.0239j,\\
n_y &= 1+0.0141j,
\end{align}
\end{subequations}
indicating that this metasurface induces both gain and loss depending on the polarization of the incident wave. Note that if we had selected bi-isotropic parameters, i.e. $\chi_\text{em}^{uu}$ and $\chi_\text{me}^{uu}$ with $u \in \{ x,y\}$, the resulting metasurface could have been lossless and purely passive (chirality).

In the non-reciprocal scenario, the metasurface is a Faraday rotating surface~\cite{Kodera_APL_07_2011,Sounas_APL_01_2011}. When the wave is sent back along the negative $z$-direction, it keeps rotating in the same absolute direction, dictated by an external biasing quantity (e.g. magnetic field or current), which doubles the total rotation (here to $\pi/8+2\times\pi/3=19\pi/24$ with respect to $x$). Therefore, reciprocal relations~\eqref{eq:chi_diag} are not appropriate and must be replaced by their counterparts obtained by selecting the non-diagonal (instead of the diagonal) terms in relations~\eqref{eq:InvProb}. This results in the non-reciprocal counterparts relations of~\eqref{eq:chi_diag}
\begin{subequations}
\label{eq:chi_off_diag}
\begin{align}
\chi_{\text{ee}}^{xy}&=\frac{-\Delta H_{y}}{j\omega\epsilon  E_{y,\text{av}}},\\
\chi_{\text{ee}}^{yx}&=\frac{\Delta H_{x}}{j\omega\epsilon  E_{x,\text{av}}},\\
\chi_{\text{mm}}^{xy}&=\frac{\Delta E_{y}}{j\omega\mu  H_{y,\text{av}}},\\
\chi_{\text{mm}}^{yx}&=\frac{-\Delta E_{x}}{j\omega\mu  H_{x,\text{av}}},
\end{align}
\end{subequations}
from which the sought susceptibilities are found to be
\begin{subequations}
\label{eq:pol_rot_imag_chi}
\begin{align}
\chi_{\text{ee}}^{xy}&=\chi_{\text{mm}}^{xy}=-0.0184j,\\
\chi_{\text{ee}}^{yx}&=\chi_{\text{mm}}^{yx}=0.0184j.
\end{align}
\end{subequations}
As expected from the prescribed non-reciprocity, these relations do not satisfy the reciprocity conditions~\eqref{eq:reciprocity}. Moreover, the sign difference between the non-diagonal elements indicates Faraday rotation~\cite{kong1986electromagnetic}. In this case and contrary to the above reciprocal example, even though all the susceptibility components are imaginary, the final metasurface is passive and lossless, as may be seen by computing the corresponding refractive indices,
\begin{subequations}
\label{eq:real_indices}
\begin{align}
n_{xy} &= \sqrt{\chi_{\text{ee}}^{xy}\chi_{\text{mm}}^{yx}}=0.0184,\\
n_{yx} &= \sqrt{\chi_{\text{ee}}^{yx}\chi_{\text{mm}}^{xy}}=0.0184.
\end{align}
\end{subequations}

\subsection{Bessel Vortex Beam Generation}
\label{ex:orbital}

\noindent \emph{Problem:} Synthesize a metasurface transforming a normally incident plane wave into a normally transmitted Bessel wave with orbital angular momentum $n=+3$.

\begin{figure}[h!]
\subfloat[]{\includegraphics[width=0.45\linewidth]{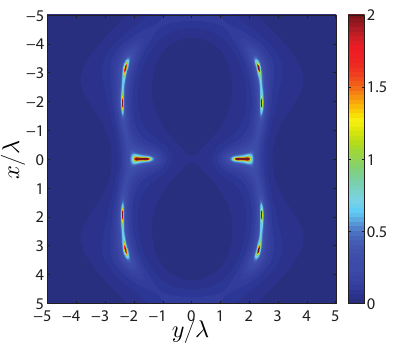}}
\subfloat[]{\includegraphics[width=0.448\linewidth]{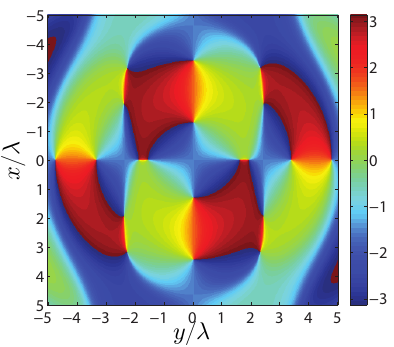}}\\
\subfloat[]{\includegraphics[width=0.45\linewidth]{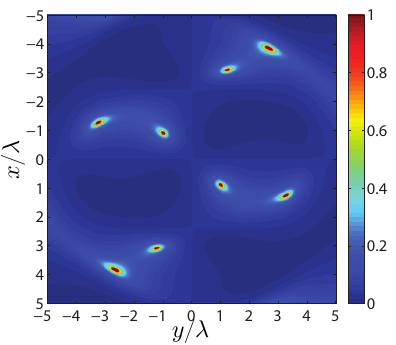}}
\subfloat[]{\includegraphics[width=0.448\linewidth]{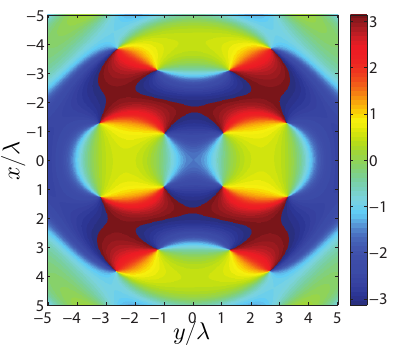}}
\caption{Magnitude (a) and phase (b) of $\chi_{\text{ee}}^{xx}$, and magnitude (c) and phase (d) of $\chi_{\text{ee}}^{yy}$ for transformation of a normally incident plane wave into a normally transmitted Bessel wave carrying orbital angular momentum with topological charge~$n=+3$.}\label{fig:bessel_example}
\end{figure}

\begin{figure}[h!]
\subfloat[]{\includegraphics[width=0.45\linewidth]{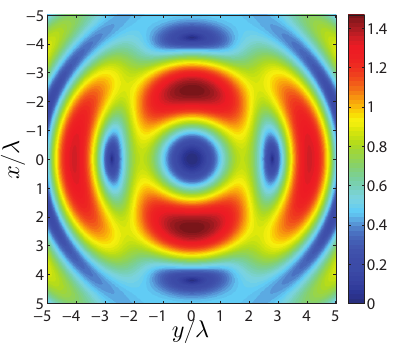}\label{fig:TRcoef_bess_a}}
\subfloat[]{\includegraphics[width=0.442\linewidth]{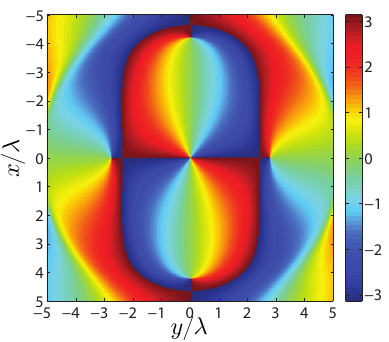}\label{fig:TRcoef_bess_b}}
\caption{Magnitude (a) and phase (b) of $T_x$ for the problem of Fig.~\ref{fig:bessel_example}.}
\label{fig:TRcoef_bess}
\end{figure}

\noindent \emph{Synthesis:} The prescribed incident and transmitted fields at $z=0$ read
\begin{subequations}
\label{eq:eq_PW_orbital}
\begin{align}
\ve{E}^{\text{i}}(x,y)&=(\ve{\hat{x}}+\ve{\hat{y}})\frac{\sqrt{2}}{2},\\
\ve{H}^{\text{i}}(x,y)&=(\ve{\hat{y}}-\ve{\hat{x}})\frac{\sqrt{2}}{2\eta},
\end{align}
\end{subequations}
and~\cite{SalemXwave}
\begin{subequations}
\label{eq:Bessel}
\begin{align}
E^{\text{t}}_\phi(\rho,\phi)&=
\frac{e^{jn\phi}}{k_\rho^2}(jA_ek_\rho\mu\omega\frac{\partial}{\partial{\rho}}
+A_h\frac{k_z n}{\rho})J_n(k_\rho\rho),\\
E^{\text{t}}_\rho(\rho,\phi)&=
\frac{e^{jn\phi}}{k_\rho^2}(A_e\frac{n\mu\omega}{\rho}
-jA_hk_\rho k_z\frac{\partial}{\partial{\rho}})J_n(k_\rho\rho),\\
E^{\text{t}}_z(\rho,\phi)&=A_hJ_n(k_\rho\rho)e^{jn\phi},\\
H^{\text{t}}_\phi(\rho,\phi)&=
\frac{e^{jn\phi}}{k_\rho^2}(A_e\frac{n k_z}{\rho}
-jA_hk_\rho\epsilon\omega\frac{\partial}{\partial{\rho}})J_n(k_\rho\rho),\\
H^{\text{t}}_\rho(\rho,\phi)&=\frac{e^{jn\phi}}{k_\rho^2}(-jA_ek_\rho k_z\frac{\partial}{\partial{\rho}}-A_h\frac{n \epsilon\omega}{\rho})J_n(k_\rho\rho),\\
H^{\text{t}}_z(\rho,\phi)&=A_eJ_n(k_\rho\rho)e^{jn\phi},
\end{align}
\end{subequations}
where the latter fields have been written in cylindrical coordinates for compactness and for a general topological charge $n$, and where $A_e$ and $A_h$ are the complex amplitudes for $e$ (TM$_z$) or $h$ (TE$_z$) polarizations, respectively. The wavevector components $k_z$ and $k_\rho$ are given in terms of the ``cone angle'' (angle between the plane waves forming the beam and the $z$ axis), $\xi$, as \mbox{$k_z=k\cos\xi$} and $k_\rho=k\sin\xi$. Here, $\xi$ is chosen such that $k_z=4k_\rho$. The Bessel wave is assumed here to be TE$_z$-polarized (i.e. $A_h=0$), where the value of $A_e$ is defined by equating the Poynting vectors ($\ve{S}=\frac{1}{2}\text{Re}\{\ve{E}\times\ve{H}^\ast\}$) of the incident and transmitted fields over the area of the metasurface (here, the surface is assumed to be $10\lambda\times10\lambda$ in size), namely $\mbox{$\iint_S \text{Re}\{\ve{E}^{\text{i}}\times \ve{H}^{\text{i}\ast}\} dS = \iint_S \text{Re}\{\ve{E}^{\text{t}}\times \ve{H}^{\text{t}\ast}\} dS$}$. This ensures that the power is conserved at the interface. Figure~\ref{fig:bessel_example} shows the metasurface susceptibilities $\chi_{\text{ee}}^{xx}$ and $\chi_{\text{ee}}^{yy}$. The field expressions in~\eqref{eq:Bessel} indicate that both the $x$ and $y$ components of the electric and magnetic fields are non-zero. Therefore, the incident fields must have electric and magnetic components along both the $x$ and $y$ directions, as chosen in~\eqref{eq:eq_PW_orbital}. In other words, the incident field must be properly polarized with respect to the prescribed transmitted wave [here~Eq.~\eqref{eq:Bessel}].
Figure~\ref{fig:TRcoef_bess} plots the $x$-components of the transmission parameter. The $y$-components (not shown) are similar to the $x$ components and rotated by $90^{\circ}$ with respect to them.

\begin{figure}[h!]
\centering
\includegraphics[width=0.52\linewidth]{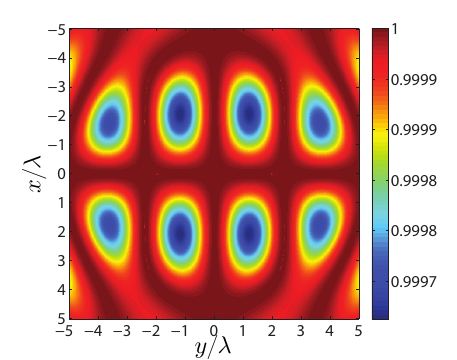}
\caption{Magnitude of the approximate transmission coefficient $T_{x,\text{approx}}$ for the problem of Fig.~\ref{fig:bessel_example}.}
\label{fig:TRcoef_bess_approx}
\end{figure}

One next applies the optimization procedure described in Sec.~\ref{sec:single_transf} to find the approximate complex transmission and reflection coefficients corresponding to a purely passive and lossless metasurface. Figure~\ref{fig:TRcoef_bess_approx} shows the corresponding transmission coefficient magnitude. Although the pattern is nonuniform, its dynamic range is close to zero with $T\approx1$ while the phase of the transmission is the same as in Fig.~\ref{fig:TRcoef_bess_b}. The magnitude null observed at the center of the $n=+3$ beam in Fig.~\ref{fig:TRcoef_bess_a} is consistent with the fact that a Bessel beam of nonzero order must exhibit an intensity null at its center due to its phase singularity at this point. In this example, this null is due to absorption since the reflected wave is specified to be zero. However, the magnitude of the approximate transmission coefficient in Fig.~\ref{fig:TRcoef_bess_approx} is almost one everywhere. In this scenario, when the approximate transmission and reflection coefficients are considered, the central null in the intensity of the transmitted Bessel beam is achieved not through absorption but through destructive interferences due to azimuthal phase variation about the center of the metasurface.
%
\subsection{Orbital Angular Momentum Multiplexing}

\noindent \emph{Problem:} Synthesize a metasurface independently transforming two normally incident plane waves with orthogonal polarizations into two normally incident transmitted Bessel waves of opposite topological charges, $n=+3$ and $n=-3$.

\begin{figure}[h!]
\subfloat[]{\includegraphics[width=0.45\linewidth]{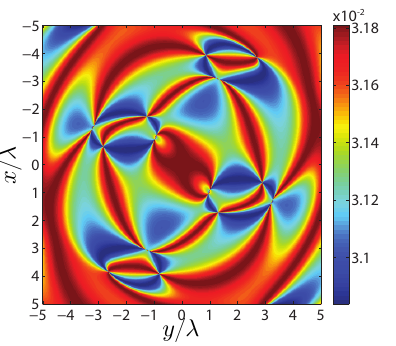}}
\subfloat[]{\includegraphics[width=0.448\linewidth]{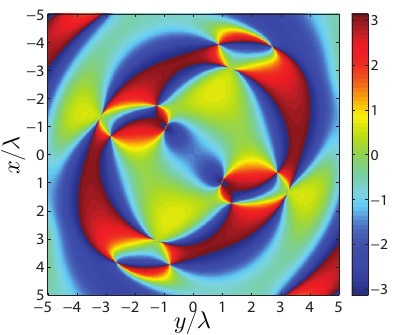}}
\caption{Magnitude (a) and phase (b) of $\chi_{\text{ee}}^{xx}$ for a transformation of two normally incident plane waves with orthogonal polarization into two normally transmitted Bessel waves carrying orbital angular momentum with topological charges $n=+3$ and $n=-3$, respectively.}\label{fig:multi_bessel_example}
\end{figure}

\noindent \emph{Synthesis:} This problem is very similar to that of Sec.~\ref{ex:orbital}, except that two waves are prescribed instead of one ($T=2$ instead of $T=1$ in Sec.~\ref{sec:synthesis}). Hence, the number of surface susceptibility components is $8$ instead of $4$, and Eq.~\eqref{eq:fullC} is to be used instead of Eq.~\eqref{eq:chi_diag}. The electromagnetic fields for the first set of incident and transmitted waves are the same as those in Eqs.~\eqref{eq:eq_PW_orbital} and~\eqref{eq:Bessel}. For the second set, the polarization of the incident plane wave is rotated by $\pi/2$ clockwise with respect to the first incident plane wave, i.e.
\begin{subequations}
\begin{align}
\ve{E}^{\text{i}}_2(x,y)&=(\ve{\hat{y}}-\ve{\hat{x}})\frac{\sqrt{2}}{2},\\
\ve{H}^{\text{i}}_2(x,y)&=-(\ve{\hat{y}}+\ve{\hat{x}})\frac{\sqrt{2}}{2\eta}.
\end{align}
\end{subequations}
The expressions for both transmitted Bessel waves remain the same, as in~\eqref{eq:Bessel}, with $n=+3$ and $n=-3$.
As in Sec.~\ref{ex:orbital}, the Bessel waves are assumed to  be TE$_z$-polarized and the values of $A_{\text{e1}}$ and $A_{\text{e2}}$ are defined by equating the Poynting vectors on both sides of the metasurface for the two cases. Figure~\ref{fig:multi_bessel_example} shows the susceptibility component $\chi_{\text{ee}}^{xx}$, as an illustration of one of the $8$ components given in~\eqref{eq:fullC}.
In this double-wave transformation, the metasurface operates as an orbital angular momentum (OAM) multiplexer, where two independent and orthogonal waves are transmitted in the same direction, with the same frequency, time and polarization, using two different OAM channels. Given its fundamental quadruple wave transformation capability (Sec.~\ref{sec:gen_sol}), the metasurface could operate, using its 16~transverse susceptibility components, as a 4 OAM-channel multiplexer.

\section{Conclusion}
\label{sec:conc}

A metasurface synthesis method based on transverse susceptibility tensors has been introduced. The technique provides closed-form expressions for selected electric and magnetic susceptibility components to theoretically treat electromagnetic transformations where the incident, reflected and transmitted waves can be specified arbitrarily. The metasurface can be reflection-less or transmission-less and can have an infinite or a finite size. Moreover, it has been shown that, by selecting more transverse susceptibility components, it is possible to treat several (up to four) sets of independent electromagnetic transformations with the same metasurface, thus allowing multi-functionality. As illustrated by the presented examples, the proposed method can handle, among others, reciprocal or non-reciprocal electromagnetic transformations, generalized refraction, polarization rotation, Bessel vortex beam generation and orbital angular momentum multiplexing. In other words, the method can be used to perform any electromagnetic transformation, without needing to resort to case-specific synthesis techniques.

The proposed synthesis method is essentially theoretical at this stage, and the physical scattering particles that would correspond to the synthesized ideal susceptibilities might be, in some cases, practically difficult or even impossible to realize. However, even in relatively extreme cases, typically corresponding to fast susceptibility variations in comparison with the wavelength, the proposed synthesis might be used as an initial and insightful step of the complete synthesis. Moreover, if required, one could practically relax some of the used assumptions, including the zero thickness of the metasurface, introducing non-zero longitudinal dipole moments ($P_z$ and $M_z$), and allowing more non-zero susceptibility tensor components. In all cases, more research is required to further develop the method and to apply it to real metasurfaces. Future possible developments may include extension to multiple layers and polychromatic waves.

\appendices
\section{Distribution Based Generalized Sheet Transition Conditions (GSTCs)~\cite{Idemen1973,kuester2003av}}
\label{Appendix}

A function $f(z)$ that is discontinuous up to the $N^\text{th}$ order at $z=0$ may be expressed in the sense of distributions as
\begin{equation}
\label{eq:General_F}
f(z)=\left \{ f(z) \right\}+\sum_{k=0}^{N}f_{k}\delta^{(k)}(z).
\end{equation}
\noindent In this relation, $\left\{f(z)\right\}$ and $\sum_{k=0}^{N}f_{k}\delta^{(k)}(z)$ are the regular and singular parts of $f(z)$, respectively. The regular part is defined for $z\neq0$ in the sense of usual functions as
\begin{equation}
\label{eq:F}
\left\{f(z)\right\}=f_{+}(z)U(z) + f_{-}(z)U(-z),
\end{equation}
\noindent where $U(x)$ is the unit step function and $f_{\pm}(z)$ denote the parts of $f(z)$ in the regions $z\gtrless{0}$. The singular part, defined at $z=0$, is a Taylor-type series, where $\delta^{(k)}(z)$ is the $k^\text{th}$ derivative of the Dirac delta function, and $f_k$ is the corresponding weighting coefficient, which is $z$-independent.

The function $f(z)$ in \eqref{eq:General_F} represents here any of the quantities in Maxwell equations. Since these equations involve spatial derivatives, the question arises as how to compute the $z$-derivative of $f(z)$. Since $f_k$ does not depend on $z$, taking the $z$-derivatives of the singular part of~\eqref{eq:General_F} only increases the derivative order of the Dirac delta function, from $k$ to $k+1$. On the other hand, the derivative of the regular part, given by~\eqref{eq:F}, involves the derivative of $U(\cdot)$, which may be expressed in the sense of distributions, in connection with a test function $\phi$, as
\begin{equation}
\label{eq:Derv_U}
\left \langle U',\phi  \right \rangle=-\left \langle U,\phi'  \right \rangle=\left \langle \delta ,\phi  \right \rangle,
\end{equation}
where $\langle\cdot,\cdot\rangle$ represents the functional inner product. In \eqref{eq:Derv_U}, the first equality was obtained by integrating by part and taking into account the fact that $\phi$ has a finite support, while the second equality follows from setting the lower bound of the integral to zero for eliminating $U$, using the fact that the primitive of $\phi'$ is $\phi$, by definition, and again that $\phi$ has a finite support, and finally applying the sifting property of the Dirac delta function according to which $\phi(0)=\langle\delta,\phi\rangle$. In other words, the derivative of the unit step function is the Dirac delta function. Therefore, using \eqref{eq:Derv_U}, the $z$-derivative of~\eqref{eq:F} is obtained as
\begin{equation}
\label{eq:Derv_F}
\begin{split}
\frac{d}{dz}\left\{f(z)\right\}&=\left \{ f_{+}'(z)U(z) + f_{-}'(z)U(-z) \right \}\\
&\qquad+[f_{+}(0)-f_{-}(0)]\delta(z)\\
&=\left \{ f' \right \}+\left [ \left [ f \right ] \right ]\delta(z),
\end{split}
\end{equation}
\noindent where $\left \{f' \right \}$ (curl bracket term in the second equality) represents the regular part of the derivative of $f(z)$, defined at $z\neq0$, and the term $\left [ \left [ f \right ] \right ]$ (square bracket term in the second equality) represents the singularity, at $z=0$. Remember that the unit of $\delta(z)$ is $(z)^{-1}$ since $\int_{-\infty}^{+\infty}\delta(z)dz=1$ is dimensionless.

Rigorous GSTCs can now be derived using \eqref{eq:General_F} and~\eqref{eq:Derv_F}. The derivation is performed here only for Maxwell-Amp$\grave{\mathrm{e}}$re equation, as the derivations for the other Maxwell equations are essentially similar. Maxwell-Amp$\grave{\mathrm{e}}$re equation in the monochromatic regime reads
\begin{equation}
\label{eq:Max-amp}
\nabla\times\ve{H}=\ve{J}+j\omega\ve{D}.
\end{equation}

\noindent Expressing $\ve{H}$ in the form~\eqref{eq:General_F} and using the transverse-longitudinal decomposition $\nabla=\nabla_{\parallel}+\hat{z}\frac{\partial }{\partial z}$ transforms the left-hand side of~\eqref{eq:Max-amp} into
\begin{equation}
\label{eq:Curl_H}
\begin{split}
\nabla\times\ve{H}&=\nabla_{\parallel}\times\left \{ \ve{H} \right \} + \hat{z}\times\frac{\partial }{\partial z}\left \{ \ve{H} \right \}\\
&\quad+\sum_{k=0}^{N}\nabla_{\parallel}\times\ve{H}_{k}\delta^{(k)}(z)+\sum_{k=0}^{N}\hat{z}\times\frac{\partial }{\partial z}\ve{H}_{k}\delta^{(k)}(z).
\end{split}
\end{equation}
\noindent In the right-hand side of~\eqref{eq:Curl_H}, the second term can be evaluated using~\eqref{eq:Derv_F} while the derivative in the last term only affects the Dirac delta function since $\ve{H}_{k}$ does not depend on $z$. Therefore, Eq.~\eqref{eq:Curl_H} becomes
\begin{equation}
\label{eq:Curl_H2}
\begin{split}
\nabla\times\ve{H}&=\nabla_{\parallel}\times\left \{ \ve{H} \right \} + \hat{z}\times\left \{\frac{\partial }{\partial z}\ve{H} \right \}+\hat{z}\times\left [ \left [ \ve{H} \right ] \right ]\delta(z)\\
&\quad+\sum_{k=0}^{N}\nabla_{\parallel}\times\ve{H}_{k}\delta^{(k)}(z)+\sum_{k=0}^{N}\hat{z}\times\ve{H}_{k}\delta^{(k+1)}(z),
\end{split}
\end{equation}
\noindent where the first two terms and the last two terms are the regular and singular parts, respectively.

Substituting~\eqref{eq:Curl_H2} along with the~\eqref{eq:General_F} expressions of $\ve{D}$ and $\ve{J}$ into~\eqref{eq:Max-amp} finally transforms Maxwell-Amp\`{e}re equation into
\begin{equation}
\label{eq:Curl_H3}
\begin{split}
&\nabla_{\parallel}\times\left \{ \ve{H} \right \} + \hat{z}\times\left \{\frac{\partial }{\partial z}\ve{H} \right \}+\hat{z}\times\left [ \left [ \ve{H} \right ] \right ]\delta(z)\\
&\quad+\sum_{k=0}^{N}\nabla_{\parallel}\times\ve{H}_{k}\delta^{(k)}(z)+\sum_{k=0}^{N}\hat{z}\times\ve{H}_{k}\delta^{(k+1)}(z)\\
&=\left\{\mathbf{J}(z)\right\}+\sum_{k=0}^{N}\mathbf{J}_{k}\delta^{(k)}(z)
+j\omega\left\{\mathbf{D}(z)\right\}+j\omega\sum_{k=0}^{N}\mathbf{D}_{k}\delta^{(k)}(z),
\end{split}
\end{equation}
where $\left\{\ve{J}(z)\right\}$ is a volume current, measured in (A/m$^2$), while $\mathbf{J}_k$ represents surface currents, measured in (A/m) since the unit of $\delta^{(k)}(z)$ is (1/m). One may now equate the terms of the same discontinuity orders, i.e. of same Dirac derivative orders, in this equation and in the other three corresponding Maxwell equations\footnote{Rigorously, the Dirac delta function disappears upon integrating over $z$ terms of equal discontinuity order.}. The result is, for the terms of order $\delta^{(0)}(z)=\delta(z)$,
\begin{subequations}\label{eq:Maxwell_distr_form}
\begin{equation}\label{eq:Maxwell_distr_form1}
\hat{z}\times\left [ \left [ \ve{H} \right ] \right ]+\nabla_{\parallel}\times\ve{H}_0
=\ve{J}_0+j\omega\ve{D}_0,
\end{equation}
\begin{equation}
\hat{z}\times\left [ \left [ \ve{E} \right ] \right ]+\nabla_{\parallel}\times\ve{E}_0
=-\ve{K}_0-j\omega\ve{B}_0,\\
\end{equation}
\begin{equation}
\hat{z}\cdot\left [ \left [ \ve{D} \right ] \right ]+\nabla_{\parallel}\cdot\ve{D}_0
=\rho_0,\\
\end{equation}
\begin{equation}
\hat{z}\cdot\left [ \left [ \ve{B} \right ] \right ]+\nabla_{\parallel}\cdot\ve{B}_0
=m_0,
\end{equation}
\end{subequations}
and, for the terms of order $\delta^{(k)}(z)$ with $k\geq1$,
\begin{subequations}\label{eq:Maxwell_distr_compat}
\begin{equation}
\hat{z}\times\ve{H}_{k-1}+\nabla_{\parallel}\times\ve{H}_k
=\ve{J}_k+j\omega\ve{D}_k,
\end{equation}
\begin{equation}
\hat{z}\times\ve{E}_{k-1}+\nabla_{\parallel}\times\ve{E}_k
=-\ve{K}_k-j\omega\ve{B}_k,\\
\end{equation}
\begin{equation}
\hat{z}\cdot\ve{D}_{k-1}+\nabla_{\parallel}\cdot\ve{D}_k
=\rho_k,\\
\end{equation}
\begin{equation}
\hat{z}\cdot\ve{B}_{k-1}+\nabla_{\parallel}\cdot\ve{B}_k
=m_k.
\end{equation}
\end{subequations}
Equations~\eqref{eq:Maxwell_distr_form} are the \emph{universal boundary conditions for monochromatic waves at a planar surface at rest}, while Eqs.~\eqref{eq:Maxwell_distr_compat} are \emph{compatibility relations} that must to be recursively applied to determine the unknown terms in~\eqref{eq:Maxwell_distr_form}~\cite{Idemen1973}. Note, letting $z\rightarrow{0}$ in the regular parts of~\eqref{eq:Maxwell_distr_form}, the presence of additional terms compared to the case of conventional boundary conditions (e.g.~Eq.~\eqref{eq:Maxwell_distr_form1}, where $\left[\left[\ve{H}\right]\right]=\left[H(z=0_+)-H(z=0_-)\right]$ and $\mathbf{J}_0$ is the sheet surface current, includes the additional terms $\nabla_\parallel\times\mathbf{H}_0$ and $j\omega\mathbf{D}_0$).

Let us now specialize to the case of interest: an \emph{infinitesimal sheet discontinuity in free space}. This means that the quantities $\ve{J}_k$, $\ve{K}_k$, $\rho_k$ and $m_k$ exclusively reside at $z=0$, so that $\ve{J}_k\equiv\ve{K}_k\equiv\rho_k\equiv m_k\equiv{0}$ for $k\geq1$, meaning that only the term $k=0$ survives in the series~\eqref{eq:General_F} for these quantities. However, the situation is different for the fields $\ve{E}_k$, $\ve{E}_k$, $\ve{D}_k$ and $\ve{B}_k$, since these fields exist also at $z\neq{0}$. Strictly, $N\rightarrow\infty$ for these fields. However, since the discontinuity is purely concentrated at $z=0$, the Taylor-type series in~\eqref{eq:General_F} includes only a small number of significant terms, and the series can be safely truncated at some value of $N$. Choosing some value for $N$ (e.g. $N=2$), Eqs.~\eqref{eq:Maxwell_distr_compat} may be solved recursively for $k=N$ to $k=1$, with $\ve{D}_k=\epsilon \ve{E}_k$ and $\ve{B}_k=\mu \ve{H}_k$. This procedure reduces the compatibility relations to
\begin{subequations}
\label{eq:comp_rel_0}
\begin{align}
\hat{z}\times\ve{H}_{0}&=0,\\
\hat{z}\times\ve{E}_{0}&=0,\\
\hat{z}\cdot\ve{D}_{0}&=0,\\
\hat{z}\cdot\ve{B}_{0}&=0.
\end{align}
\end{subequations}

One may now introduce the electric and magnetic polarization densities, $\mathbf{P}$ and $\mathbf{M}$, respectively, to account for the action of the scattering particles forming the metasurface. For this purpose, the standard constitutive relations $\ve{D}=\epsilon \ve{E}+\ve{P}$ and $\ve{B}=\mu (\ve{H}+\ve{M})$ are in a form that properly models the first order surface discontinuity in \eqref{eq:Maxwell_distr_form}, namely
\begin{subequations}\label{eq:DHs}
\begin{equation}
\ve{D}_0=\epsilon \ve{E}_0+\ve{P}_0,
\end{equation}
\begin{equation}
\ve{H}_0=\frac{1}{\mu }\ve{B}_0-\ve{M}_0,
\end{equation}
\end{subequations}
where $\ve{P}_0$ and $\ve{M}_0$ represent the (first-order) electric and magnetic \emph{surface} polarization densities, respectively. In the absence of sources ($\ve{J}_0=\ve{K}_0=\rho_0=m_0=0$), substitution of~\eqref{eq:DHs} and application of \eqref{eq:comp_rel_0} transforms \eqref{eq:Maxwell_distr_form} into
\begin{subequations}\label{eq:final_BC}
\begin{equation}
\hat{z}\times\left [ \left [ \ve{H} \right ] \right ]
=j\omega\ve{D}_0-\nabla_\parallel\times\ve{H}_0
=j\omega\ve{P}_{0,\parallel}+\nabla_{\parallel}\times\ve{M}_{0,\text{n}},
\end{equation}
\begin{equation}
\hat{z}\times\left [ \left [ \ve{E} \right ] \right ]
=-j\omega\ve{B}_0-\nabla_\parallel\times\ve{E}_0
=-j\omega\mu \ve{M}_{0,\parallel}+\frac{1}{\epsilon }\nabla_{\parallel}\times\ve{P}_{0,\text{n}},
\end{equation}
\begin{equation}
\hat{z}\cdot\left [ \left [ \ve{D} \right ] \right ]
=-\nabla_\parallel\cdot\ve{D}_0
=-\nabla_{\parallel}\cdot\ve{P}_{0,\parallel},
\end{equation}
\begin{equation}
\hat{z}\cdot\left [ \left [ \ve{B} \right ] \right ]
=-\nabla_\parallel\cdot\ve{B}_0
=-\mu \nabla_{\parallel}\cdot\ve{M}_{0,\parallel},
\end{equation}
\end{subequations}
where the subscripts $\parallel$ and n denote transverse and normal components, respectively.

Using the relation $\nabla_\parallel\times(\ve{\hat{z}}\psi)=-\ve{\hat{z}}\times\nabla_\parallel\psi$
and the difference notation~\eqref{eq:field_diff}, Eqs.~\eqref{eq:final_BC} finally take the form
\begin{subequations}
\label{eq:BCfinapp}
\begin{align}
\hat{z}\times\Delta\ve{H}
&=j\omega\ve{P}_{\parallel}-\hat{z}\times\nabla_{\parallel}M_{z},\\
\Delta\ve{E}\times\hat{z}
&=j\omega\mu \ve{M}_{\parallel}-\nabla_{\parallel}\bigg(\frac{P_{z}}{\epsilon }\bigg)\times\hat{z},\\
\hat{z}\cdot\Delta\ve{D}
&=-\nabla\cdot\ve{P}_{\parallel},\\
\hat{z}\cdot\Delta\ve{B}
&=-\mu \nabla\cdot\ve{M}_{\parallel}.
\end{align}
\end{subequations}

It has been implicitly assumed throughout these derivations that the metasurface is surrounded by two identical media, with permittivity and permeability $\epsilon$ and $\mu$, respectively. If the two media are different, then the last two equations of~\eqref{eq:BCfinapp} contain the different media parameters. For instance, in the third equation, one would have $\Delta\ve{D}=\ve{D}^\text{t}-(\ve{D}^\text{i}+\ve{D}^\text{r})=\epsilon_+\ve{E}^\text{t}-\epsilon_-(\ve{E}^\text{i}-\ve{E}^\text{r})$.

\bibliographystyle{IEEEtran}
\bibliography{TAP_METASURFACE_SUSCEPT_SYNTHESIS_Achouri}

\begin{thebibliography}{10}
\providecommand{\url}[1]{#1}
\csname url@samestyle\endcsname
\providecommand{\newblock}{\relax}
\providecommand{\bibinfo}[2]{#2}
\providecommand{\BIBentrySTDinterwordspacing}{\spaceskip=0pt\relax}
\providecommand{\BIBentryALTinterwordstretchfactor}{4}
\providecommand{\BIBentryALTinterwordspacing}{\spaceskip=\fontdimen2\font plus
\BIBentryALTinterwordstretchfactor\fontdimen3\font minus
  \fontdimen4\font\relax}
\providecommand{\BIBforeignlanguage}[2]{{%
\expandafter\ifx\csname l@#1\endcsname\relax
\typeout{** WARNING: IEEEtran.bst: No hyphenation pattern has been}%
\typeout{** loaded for the language `#1'. Using the pattern for}%
\typeout{** the default language instead.}%
\else
\language=\csname l@#1\endcsname
\fi
#2}}
\providecommand{\BIBdecl}{\relax}
\BIBdecl

\bibitem{kuester2003av}
E.~F. Kuester, M.~Mohamed, M.~Piket-May, and C.~Holloway, ``Averaged transition
  conditions for electromagnetic fields at a metafilm,'' \emph{IEEE Trans.
  Antennas Propag.}, vol.~51, no.~10, pp. 2641--2651, Oct 2003.

\bibitem{Holloway2009}
C.~Holloway, A.~Dienstfrey, E.~F. Kuester, J.~F. O'Hara, A.~K. Azad, and A.~J.
  Taylor, ``{A discussion on the interpretation and characterization of
  metafilms/metasurfaces: the two-dimensional equivalent of metamaterials},''
  \emph{Metamaterials}, vol.~3, no.~2, pp. 100--112, Oct. 2009.

\bibitem{holloway2012overview}
C.~Holloway, E.~F. Kuester, J.~Gordon, J.~O'Hara, J.~Booth, and D.~Smith, ``An
  overview of the theory and applications of metasurfaces: the two-dimensional
  equivalents of metamaterials,'' \emph{IEEE Antennas Propag. Mag.}, vol.~54,
  no.~2, pp. 10--35, April 2012.

\bibitem{capolino2009theory}
F.~Capolino, \emph{Theory and phenomena of metamaterials}.\hskip 1em plus 0.5em
  minus 0.4em\relax CRC Press, 2009.

\bibitem{engheta2006metamaterials}
N.~Engheta and R.~W. Ziolkowski, \emph{Metamaterials: physics and engineering
  explorations}.\hskip 1em plus 0.5em minus 0.4em\relax John Wiley \& Sons,
  2006.

\bibitem{caloz2005electromagnetic}
C.~Caloz and T.~Itoh, \emph{Electromagnetic metamaterials: transmission line
  theory and microwave applications}.\hskip 1em plus 0.5em minus 0.4em\relax
  John Wiley \& Sons, 2005.

\bibitem{MunkFSS}
B.~A. Munk, \emph{Frequency Selective Surfaces: theory and design}.\hskip 1em
  plus 0.5em minus 0.4em\relax John Wiley \& Sons, 2000.

\bibitem{holloway2005reflection}
C.~Holloway, M.~Mohamed, E.~F. Kuester, and A.~Dienstfrey, ``Reflection and
  transmission properties of a metafilm: with an application to a controllable
  surface composed of resonant particles,'' \emph{IEEE Trans. Electromagn.
  Compat.}, vol.~47, no.~4, pp. 853--865, Nov 2005.

\bibitem{Pfeiffer2013a}
C.~Pfeiffer and A.~Grbic, ``Metamaterial {H}uygens' surfaces: tailoring wave
  fronts with reflectionless sheets,'' \emph{Phys. Rev. Lett.}, vol. 110, p.
  197401, May 2013.

\bibitem{Ra2013}
Y.~Ra'di, V.~Asadchy, and S.~Tretyakov, ``Total absorption of electromagnetic
  waves in ultimately thin layers,'' \emph{IEEE Trans. Antennas Propag.},
  vol.~61, no.~9, pp. 4606--4614, Sept 2013.

\bibitem{shi2014dual}
H.~Shi, A.~Zhang, S.~Zheng, J.~Li, and Y.~Jiang, ``Dual-band polarization angle
  independent 90° polarization rotator using twisted electric-field-coupled
  resonators,'' \emph{Appl. Phys. Lett.}, vol. 104, no.~3, pp.~--, 2014.

\bibitem{capasso1}
N.~Yu, P.~Genevet, M.~A. Kats, F.~Aieta, J.-P. Tetienne, F.~Capasso, and
  Z.~Gaburro, ``Light propagation with phase discontinuities: generalized laws
  of reflection and refraction,'' \emph{Science}, vol. 334, no. 6054, pp.
  333--337, 2011.

\bibitem{PhysRevApplied.2.044011}
C.~Pfeiffer and A.~Grbic, ``Bianisotropic metasurfaces for optimal polarization
  control: Analysis and synthesis,'' \emph{Phys. Rev. Applied}, vol.~2, p.
  044011, Oct 2014.

\bibitem{6905746}
M.~Selvanayagam and G.~Eleftheriades, ``Polarization control using tensor
  huygens surfaces,'' \emph{IEEE Trans. Antennas Propag.}, vol.~62, no.~12, pp.
  6155--6168, Dec 2014.

\bibitem{6477089}
T.~Niemi, A.~Karilainen, and S.~Tretyakov, ``Synthesis of polarization
  transformers,'' \emph{IEEE Trans. Antennas Propag.}, vol.~61, no.~6, pp.
  3102--3111, June 2013.

\bibitem{Salem2013c}
M.~A. Salem and C.~Caloz, ``Manipulating light at distance by a metasurface
  using momentum transformation,'' \emph{Opt. Express}, vol.~22, no.~12, pp.
  14\,530--14\,543, Jun 2014.

\bibitem{Schelkunoff1}
S.~Schelkunoff, ``On teaching the undergraduate electromagnetic theory,''
  \emph{IEEE Trans. Educ.}, vol.~15, no.~1, pp. 15--25, Feb 1972.

\bibitem{GrbicLightBending}
C.~Pfeiffer, N.~K. Emani, A.~M. Shaltout, A.~Boltasseva, V.~M. Shalaev, and
  A.~Grbic, ``Efficient light bending with isotropic metamaterial {H}uygens’
  surfaces,'' \emph{Nano Letters}, vol.~14, no.~5, pp. 2491--2497, 2014.

\bibitem{Idemen1973}
M.~M. Idemen, \emph{Discontinuities in the Electromagnetic Field}.\hskip 1em
  plus 0.5em minus 0.4em\relax John Wiley \& Sons, 2011.

\bibitem{kong1986electromagnetic}
J.~Kong, \emph{Electromagnetic wave theory}, ser. A Wiley-Interscience
  publication.\hskip 1em plus 0.5em minus 0.4em\relax John Wiley \& Sons, 1986.

\bibitem{lindell1994electromagnetic}
I.~Lindell, \emph{Electromagnetic waves in chiral and bi-isotropic media}, ser.
  The Artech House Antenna Library.\hskip 1em plus 0.5em minus 0.4em\relax
  Artech House, 1994.

\bibitem{asadchy2011simulation}
V.~S. Asadchy and I.~A. Fanyaev, ``Simulation of the electromagnetic properties
  of helices with optimal shape, which provides radiation of a circularly
  polarized wave,'' \emph{Journal of Advanced Research in Physics}, vol.~2,
  no.~1, 2011.

\bibitem{asadchy2014determining}
V.~S. Asadchy, I.~A. Faniayeu, Y.~Ra'di, and S.~A. Tretyakov, ``Determining
  polarizability tensors for an arbitrary small electromagnetic scatterer,''
  \emph{arXiv preprint arXiv:1401.4930}, 2014.

\bibitem{Kodera_APL_07_2011}
T.~Kodera, D.~L. Sounas, and C.~Caloz, ``Artificial {F}araday rotation using a
  ring metamaterial structure without static magnetic field,'' \emph{App. Phys.
  Lett.}, vol.~99, no.~3, pp. 031\,114:1--3, Jul. 2011.

\bibitem{Sounas_APL_01_2011}
D.~L. Sounas and C.~Caloz, ``Electromagnetic non-reciprocity and gyrotropy of
  graphene,'' \emph{App. Phys. Lett.}, vol.~98, no.~2, pp. 021\,911:1--3, Jan.
  2011.

\bibitem{SalemXwave}
M.~A. Salem and H.~Ba\u{g}c{i}, ``Reflection and transmission of normally
  incident full-vector {X}-waves on planar interfaces,'' \emph{J. Opt. Soc. Am.
  A}, vol.~29, no.~1, pp. 139--152, Jan 2012.

\end{thebibliography}

\end{document}